\documentclass[11pt, a4paper]{article}
\pdfoutput=1
\usepackage[pdfauthor={P. P. Cook},
pdftitle={Bound States of String Theory and Beyond}]{hyperref}
\usepackage{amsmath}
\usepackage{graphicx}
\usepackage{youngtab}
\usepackage{ytableau}
\usepackage{amssymb}
\usepackage{amsthm}
\usepackage{subfig}
\usepackage{chngcntr}
\usepackage{array}
\usepackage{xy}
\counterwithin{figure}{section}
\counterwithin{table}{section}
\newcommand{\comment}[1]{}

\begin{document}
\title{Bound States of String Theory and Beyond}
\author{Paul P. Cook\footnote{email: paul.cook@kcl.ac.uk} \\ 
\\
{\itshape Department of Mathematics, King's College London \\ 
The Strand, London WC2R 2LS, UK}
}
\begin{titlepage}
\begin{flushright}
{\tt KCL-MTH-11-18} \\
\end{flushright}
\vspace{70pt}
\centering{\LARGE Bound States of String Theory and Beyond}\\
\vspace{30pt}
\def\thefootnote{\fnsymbol{footnote}}
Paul P. Cook\footnote{\href{mailto:paul.cook@kcl.ac.uk}{email: paul.cook@kcl.ac.uk}}\\
\setcounter{footnote}{0}
\vspace{10pt}
{\itshape Department of Mathematics, King's College London \\ 
The Strand, London WC2R 2LS, UK}
\vspace{30pt}
\begin{abstract}
All bound states of fundamental strings, D-branes and NS-branes of string theory, both type-IIA and type-IIB, which may be described by a null geodesic motion on the coset ${\cal G/K(G)}$ where ${\cal G}$ is a group of type $A_n$, $D_n$ or $E_n$ embedded within $E_{11}$ are presented.
\end{abstract}
\vspace{80pt}
\emph{In memory of Laurent Houart.}
\end{titlepage}
\clearpage
\newpage
\begin{section}{Introduction}
The modern understanding of the relevance of Kac-Moody algebras to M-theory has developed from a number of different perspectives. It has been argued that the Kac-Moody algebra $\mathfrak{e}_{11}$ is a symmetry of the extension of eleven-dimensional supergravity relevant to M-theory \cite{West:2001p131}. It has been shown that the hyperbolic Kac-Moody algebra $\mathfrak{e}_{10}$ controls the chaotic equations of motion in the vicinity of a cosmological singularity \cite{Damour:2001p3108}. Prior to this $\mathfrak{e}_{10}$ was shown to be a symmetry of bosonic supergravity dimensionally reduced to one dimension \cite{Mizoguchi:1998p4868}. That this would be the case was anticipated some time before \cite{Julia:1981p4403}. It was also shown via reduction to one dimension that ${\cal N}=1$ supergravity possessed a symmetry encoded in a hyperbolic Kac-Moody algebra \cite{Nicolai:1992p4869}. However the full purpose of Kac-Moody algebras in defining the physical theory remains unclear. One unanswered question concerns whether the tensor fields which parameterise the symmetry are all on the same footing in the related physical theory. One would assume so, however there are a distinguished set of tensor fields, the form fields, appearing as coefficients in the algebras $\mathfrak{e}_{11}$ or $\mathfrak{e}_{10}$ which are completely antisymmetrised and directly related to the gauge potentials of M-theory and string theory sourcing brane solutions. What interpretation, if any, is to be given to the mixed-symmetry tensors prior to compactification? 

Certainly the mixed symmetry tensor fields of $\mathfrak{e}_{11}$ in eleven dimensions play a number of roles in lower dimensions. Most directly mixed symmetry fields which, upon dimensional reduction, give form fields that source brane solutions, notably the D6 brane of IIA string theory is sourced by a seven form field which is the dimensional reduction of the dual graviton in $\mathfrak{e}_{11}$ \cite{Kleinschmidt:2004p371}. Similarly the doublet and quadruplet of ten-form potentials, also found in \cite{Kleinschmidt:2004p371} and derived from mixed symmetry tensors in $\mathfrak{e}_{11}$, act as a source for D9 branes in IIB supergravity \cite{Bergshoeff:2005p935}. The Romans mass parameter \cite{Schnakenburg:2002p2747} and many of the fields parameterising the gauging of supergravity \cite{Riccioni:2007p609,Riccioni:2009p1980}, see also \cite{Bergshoeff:2007p1300,Wit:2008p4770,Bergshoeff:2008p4773}, are derived from the reduction of mixed symmetry fields appearing in $\mathfrak{e}_{11}$. More recently, guided by the tension of the related objects in $\mathfrak{e}_{11}$ \cite{Cook:2008p936}, mixed symmetry fields in eleven dimensions have been interpreted in ten dimensions and less as solitons and defect branes which give rise to supersymmetric branes of maximal supergravity and enlarge the Wess-Zumino term \cite{Bergshoeff:2010p4753,Bergshoeff:2011p4751,Bergshoeff:2011p4749,Bergshoeff:2011p4758,Bergshoeff:2011p4756,Bergshoeff:2011p4742,Kleinschmidt:2011p4762}.

However in this paper we are interested to pursue the idea that there is a consistent interpretation for the mixed-symmetry tensors of $\mathfrak{e}_{11}$ directly in eleven dimensions. In this context the mixed symmetry tensors of co-dimension two, which were first discussed in $\mathfrak{e}_{10}$ in \cite{Damour:2002p617} and fully classified within $\mathfrak{e}_{11}$ in \cite{Riccioni:2006p600} have been understood as the completion of the Geroch group of solutions \cite{Geroch:1971p4776,Geroch:1972p4777,Breitenlohner:1987p4778} of eleven dimensional supergravity \cite{Englert:2007p605}. It was proposed in \cite{Cook:2009p2751} that it was possible to understand all mixed symmetry fields from $E_{11}$ as a conglomerate of form fields directly in eleven dimensions and that the resulting solution, which interpolated between the form fields, should be understood as a bound state of M-branes. Shortly thereafter this proposal was extended and it was shown that the solutions were derived from a precise model \cite{Houart:2010p2928}. Historically single string and brane solutions were found encoded in the form of a general solution-generating coset representative group element \cite{West:2004p1593} and separately as the null geodesic motion of a particle on the coset $\frac{SL(2,\mathbb{R})}{SO(1,1)}$ \cite{Englert:2004p2874}. The development of bound state solutions mirrored these earlier discoveries for single branes and the results of \cite{Houart:2010p2928} showed that bound state solutions arise as the null geodesic motion of a particle on cosets of groups larger than $SL(2,\mathbb{R})$ and that the coset algebra is embedded in $\mathfrak{e}_{11}$. 
 
A complete list of bound states solutions described in this way depends upon finding which embeddings of sub-groups ${\cal G}< E_{11}$ are possible within $E_{11}$ where $SL(2,\mathbb{R})<{\cal G}< E_{11}$. Sets of positive roots of $E_{11}$, each individually associated to a single brane solution, which form the positive roots of a sub-group ${\cal G}$ may be identified with bound states of branes. The one-dimensional bound state solution is described by a null geodesic motion on the coset $\frac{\cal G}{\cal{K(G)}}$, where ${\cal K(G)}$ is a real form of the maximal compact sub-group of ${\cal G}$. The real form of the embedded sub-algebra is derived from the real form of $E_{11}$ chosen to pick out an $SO(1,10)$ local sub-algebra. The null geodesic motion on the coset is related to space-time by identifying the parameter of the coset path to a single transverse direction in the background geometry of the brane bound state. This method effectively identifies one-dimensional M-theory solutions giving precise expressions for the geometry and gauge fields of the solution.

For example the dyonic membrane \cite{Izquierdo:1995p1636}, which describes a bound state of the membrane with the fivebrane, may be described in this setting by a null geodesic motion on $\frac{SL(3,\mathbb{R})}{SO(1,2)}$ \cite{Houart:2010p2928}. The $SL(3,\mathbb{R})$ is identified with the dyonic membrane by observing that the roots of $E_{11}$ corresponding to the M2 brane ($\beta_{M2}$), the M5 brane ($\beta_{M5}$) and an S2 brane ($\beta_{S2}=\beta_{M5}-\beta_{M2}$) form the simple positive roots of $SL(3,\mathbb{R})$\footnote{as $\beta_{M2}\cdot\beta_{S2}=-1$, $\beta_{M2}^2=\beta_{S2}^2=\beta_{M5}^2=2$ and $\beta_{M5}=\beta_{M2}+\beta_{S2}$ using the inner product on the $E_{11}$ root space.}. The solution has a global $SL(3,\mathbb{R})$ symmetry which acts to permute the individual branes, while the $SO(1,2)$ symmetry acts on the constants within the harmonic functions describing the individual solutions and may be used to turn on and off the various brane charges within the composite solution. In this context one could describe the dyonic membrane as the orbit space of the membrane generated by the action of $SO(1,2)$.

The reduction of these types of solution to IIA and IIB string theory is more interesting simply because there are so many more canonical solutions of string theory than M-theory. Within the Kac-Moody framework this corresponds to many more interesting  roots associated to branes that one can use to form the simple positive roots of the sub-group ${\cal G}$ embedded within $E_{11}$. 
In this paper we will develop the case for interpreting mixed symmetry fields in ten dimensions as bound states of Dp-branes and NS-branes directly in ten dimensions. We do this by identifying all possible sub-groups $\cal G$ which may be embedded in $E_{11}$ using generators associated only to canonical string theory branes and giving solutions for a selection of those which do not involve a mixed-symmetry tensor. It will be seen that there is no algebraic difference between the bound states of only canonical branes and those which also contain mixed-symmetry tensors or exotic content. More precisely we choose a set of branes, (D0,F1,D2,D4,NS5,D6,D8) for the type-IIA string theory and (D1,F1,D3,D5,NS5,D7$_a$,D7$_b$,D9$_a$,D9$_b$)\footnote{See section 2 for the naming conventions adopted in this paper for seven and nine branes.} for type-IIB string theory, and identify all the $A_n$, $D_n$ and $E_n$ root systems that can be constructed using the real roots of $E_{11}$ associated to these branes. The number of possible root systems is vast and so is presented in its entirety in the form of a catalogue which accompanies this paper and is available on the internet with the arxiv submission for this paper. We will survey the many results, indicating previously known solutions that arise within this framework, analysing new solutions and highlighting a variety of pathological cases which also arise.

The paper is organised as follows, in section 2 we describe the root systems relevant to the IIA and IIB decompositions of $E_{11}$ as Young tableaux. In section 3 we discuss the brane intersections, those orientations of brane for which there is no binding energy. In section 4 we describe the algorithm for identifying all $A_n$, $D_n$ and $E_n$ root systems whose simple roots are associated to single brane solutions of string theory and review the results.

\end{section}
\begin{section}{String theory fields as Young tableaux from $E_{11}$}
The algebra of $E_{11}$ may be decomposed into representations of $SL(10,\mathbb{R})$ relevant to type IIA \cite{West:2001p131} and type IIB \cite{Schnakenburg:2001p2748} string theory. In this section we present these algebra decompositions using a basis for the root space that allows the direct conversion of a root of $E_{11}$ into the Young tableau for the associated generator in the algebra. This presentation will be useful for visualising bound states of branes. $E_{11}$ can be represented by an infinite set of $SL(11,\mathbb{R})$ tensors which may be rapidly generated at any level by following a number of simple algebraic rules  \cite{Cook:2009p2751}. Within the algebra generated in this way are all the possible generators of the $E_{11}$ algebra as well as a relatively small (but infinite) number of generators that are excluded from the algebra by restrictions on their multiplicity. These excluded generators are worthy of further study, in particular a fast method for their discovery would be very useful, but will play no role in the discussions of the present paper and we will be content to use the algebra of $E_{11}$ up to multiplicity for the remainder of this work. 
\begin{subsection}{Type IIA string theory}
\begin{figure}[h]
\centering
\includegraphics[scale=0.8,angle=0]{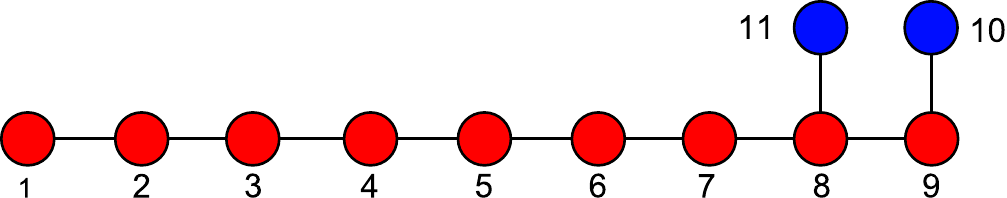}
\caption{The reduction of the Dynkin diagram for $E_{11}$} relevant to type-IIA theories \label{E11IIA}
\end{figure}
The Dynkin diagram of $E_{11}$ is shown in figure \ref{E11IIA}. One may decompose the algebra into representations of an $\mathfrak{sl}(10,\mathbb{R})$ sub-algebra whose Dynkin diagram that of figure \ref{E11IIA} once nodes ten and eleven have been deleted. Amongst the $\mathfrak{sl}(10,\mathbb{R})$ generators that arise in this decomposition are all those found in the bosonic part of IIA superstring theory. An additional, infinite number of generators are also found and these too are expected to be relevant for IIA string theory. It is the existence of these exotic generators which provides the motivation for this paper. 

Each of the deleted nodes ten and eleven indicated in figure \ref{E11IIA} give rise to a lowest weight representation of $SL(10,\mathbb{R})$. The deletion of node ten gives the $\bf{10}$, given by tensor $R^a$, and the $\bf{55}$, with tensor components $R^{a_1a_2}$. These generators are indicated by the Young tableaux:
\begin{align}
&\qquad \Yvcentermath1 \yng(1)  \qquad \qquad \qquad \qquad \qquad \qquad \qquad  \yng(1,1)  \quad \label{IIAblocks} 
\end{align}
These two Young tableaux are the building blocks of the IIA algebra containing an $E_{11}$ symmetry. Together with the generators of $\mathfrak{sl}(10,\mathbb{R})$ which we indicate by ${K^a}_b$ where $b>a$ and $a,b\in\{1,2,\ldots , 10\}$ and the Cartan sub-algebra, indicated by $H_i$ $i\in(1,2,\ldots 11)$, these form the Borel sub-algebra of $E_{11}$ relevant to the IIA theory. For reference the Cartan sub-algebra of $E_{11}$ in this decomposition is
\begin{align}
\nonumber H_i&={K^i}_i-{K^{i+1}}_{i+1} \qquad \mbox{for } i=1,2,\ldots 9,\\
H_{10}&=-\frac{1}{8}({K^1}_1+\ldots +{K^9}_9)+\frac{7}{8}{K^{10}}_{10}-\frac{3}{2}R \qquad \mbox{and}\\
\nonumber H_{11}&=-\frac{1}{4}({K^1}_1+\ldots +{K^8}_8)+\frac{3}{4}({{K^9}_9+K^{10}}_{10})+R
\end{align}
where $R$ is the generator associated to the dilaton. The Borel sub-algebra may be represented by an infinite set of $SL(10,\mathbb{R})$ tensors, of which the Young tableaux shown in (\ref{IIAblocks}) are the simplest non-trivial generators. The complete Borel sub-algebra may be visualised by stacking multiple copies of the Young tableaux in equation (\ref{IIAblocks}) while preserving  symmetry properties of these two fundamental tableau - the exact procedure for combing Young tableau in this way is encoded by the Littlewood-Richardson rules. One may combine a particular Young table of $\tiny \yng(1)$ and a Young table of $\tiny \yng(1,1)$ to form $\tiny \yng(1,1,1)$ or $\tiny \yng(2,1)$ but not $\tiny \yng(3)$.

There is a simple method that allows one to convert each highest weight Young tableau of the full algebra to a root vector. Let a generic root in the root system of $E_{11}$ be
\begin{equation}
\beta=m_1\alpha_1+m_2\alpha_2+\ldots +m_{10}\alpha_{10}+m_{11}\alpha_{11}.
\end{equation}
The decomposition of the algebra is graded by the level $(m_{10},m_{11})$, the coefficients of the deleted nodes that lead to the $\mathfrak{sl}(10,\mathbb{R})$ sub-algebra in the root $\beta$. The two Young tableaux in (\ref{IIAblocks}) are associated to the generators that appear at levels $(1,0)$ $\tiny\yng(1)$ and $(0,1)$ $\tiny\yng(1,1)$ in the grading of the decomposed algebra. The Young tableaux of the  generators at level $(m_{10},m_{11})$ therefore have $m_{10}+2m_{11}$ boxes.

In order to understand the precise shape of the Young tableaux it is useful to work with an alternative basis of the root space which directly indicates the index structure of the $SL(10,\mathbb{R})$ tensors. We take as such a basis the vectors $\{e_1,\ldots ,e_{11}\}$ in which positive simple roots of $E_{11}$ become:
\begin{align}
\nonumber \alpha_a &= e_a - e_{a+1} \qquad \mbox{(where } a\leq 10 \mbox{) and }\\
\alpha_{11} &= e_9 + e_{10} + e_{11}.\label{IIAbasis}
\end{align}
To read off a Young tableau from a root $\beta$ associated to a lowest weight representation we write the the root in terms of this $e_i$ basis: 
\begin{equation}
\beta=w_1e_1+w_2e_2+\ldots + w_{11}e_{11} \label{IIAroot}
\end{equation}
The Young tableau associated to this root $\beta$ has, reading from top to bottom, a first row of width $w_{10}$, a second row of width $w_9$ and so on down to a final row of width $w_1$. An example Young tableau for $\beta$ might have the following shape (but not the labels which simply enumerate the columns here)
\begin{equation}
\ytableausetup
{mathmode, boxsize=1.5em}
\begin{ytableau}
\mbox{{\scriptsize 1}}&\mbox{{\scriptsize 2}}&\none[\ldots]&\none[\ldots] &\none[\ldots]&\none[\ldots]&\mbox{{\scriptsize $w_{10}$}}\\
\mbox{{\scriptsize 1}}&\mbox{{\scriptsize 2}}&\none[\ldots] &\none[\ldots]&\none[\ldots]&\mbox{{\scriptsize $w_{9}$}}\\
\mbox{{\scriptsize 1}}&\mbox{{\scriptsize 2}}&\none[\ldots] &\none[\ldots]&\none[\ldots]&\mbox{{\scriptsize $w_{8}$}}\\
\none[\vdots]&\none[\vdots]&\none[\vdots]&\none[]&\none[\vdots]\\\
\mbox{{\scriptsize 1}}&\mbox{{\scriptsize 2}}&\none[\ldots]&\none[\ldots]&\mbox{{\scriptsize $w_{2}$}}\\
\mbox{{\scriptsize 1}}&\mbox{{\scriptsize 2}}&\none[\ldots] &\mbox{{\scriptsize $w_{1}$}}\\
\end{ytableau} .
\end{equation}
This root is associated to the lowest weight of an $SL(10,\mathbb{R})$ representation which guarantees that $w_1 \leq w_2 \leq \ldots \leq w_{10}$. One can see that from the roots $\alpha_{10}$ and $\alpha_{11}$ (\ref{IIAbasis}), one can rapidly find the associated Young tableau in (\ref{IIAblocks}). However given a Young tableau the inverse method to find a root is ambiguous since the coefficient $w_{11}$ in (\ref{IIAroot}) is not encoded in the Young tableau. Consequently we will label each IIA Young tableau with an additional number $\lambda\equiv w_{11}$, which encodes the second level in the decomposition, so that each Young tableau and $\lambda$ pair gives a unique root in the algebra. In fact $\lambda$ is given in terms of $m_{10}$ and $m_{11}$ by
\begin{equation}
\lambda=m_{11}-m_{10}
\end{equation}
hence the number of boxes, $\#$, in a IIA Young tableau with a given $\lambda$ is
\begin{equation}
\#=3m_{11}-\lambda.
\end{equation}
and we note that we may therefore parameterise the generators appearing in the decomposition by $(m_{11},\lambda)$ instead of $(m_{10},m_{11})$. 
\begin{table}
\centering 
\scalebox {0.95} {
\begin{tabular}{| c | c | c | c | c | c |} 
\hline Level, $m_{11}$ &  $\lambda=-1$ & $\lambda=0$ &  $\lambda=1$ &  $\lambda=2$ &  $\lambda=3$\\ [0.5ex]	
 \hline  
0 &$\tiny\yng(1)$ & $\tiny\bullet $ &&& \\ 
& D0 & Dilaton&&& \\
 \hline &&&&& \\
1 &&$\tiny\yng(1,1,1)$ & $\tiny\yng(1,1)$ && \\
&& D2 &  F1 && \\
 \hline &&&&& \\
2 &&$\tiny\yng(1,1,1,1,1,1)$ & $\tiny\yng(1,1,1,1,1)$ && \\
&& NS5 &  D4 && \\
 \hline &&&&& \\
 3&& $\tiny\yng(1,1,1,1,1,1,1,1,1)\, ,\tiny\yng(2,1,1,1,1,1,1,1)$ &$\tiny\yng(1,1,1,1,1,1,1,1)\, , \tiny\yng(2,1,1,1,1,1,1)$ &$\tiny\yng(1,1,1,1,1,1,1)$ &\\ 
 && &&D6 &\\ 
 \hline &&&&& \\
 4&& $\tiny\yng(2,2,1,1,1,1,1,1,1,1)\, ,\tiny\yng(3,1,1,1,1,1,1,1,1,1) \, ,\tiny\yng(2,2,2,1,1,1,1,1,1) $ & $\tiny\yng(2,1,1,1,1,1,1,1,1,1) \, ,\tiny\yng(2,2,1,1,1,1,1,1,1) ,\,\tiny\yng(3,1,1,1,1,1,1,1,1) , \,\tiny\yng(2,2,2,1,1,1,1,1)$ & $\tiny\yng(1,1,1,1,1,1,1,1,1,1) \, ,\tiny\yng(2,1,1,1,1,1,1,1,1) , \,\tiny\yng(2,2,1,1,1,1,1,1)$ & $\tiny\yng(1,1,1,1,1,1,1,1,1)$ \\ 
 &&&&&D8\\
 \hline 
 \end{tabular} }
 \caption{Low level IIA Young tableaux from $\mathfrak{e}_{11}$. By $\bullet$ we indicate a Young tableau for a scalar. The generators of $\mathfrak{sl}(10,\mathbb{R})$ also appear at level 0.} \label{table:IIA} 
\end{table}
We have given a method to translate (lowest weight) roots in the algebra into Young tableau and now we turn our attention to the algebraic condition which roots $\beta$ must satisfy in order to appear in the root space of $E_{11}$. The condition that a root $\beta$ exists in the root system, derived from the Serre relations, is simply that
\begin{equation}
\beta^2= 2, 0, -2, -4 \ldots
\end{equation}
where the root length squared formula is
\begin{equation}
\beta^2=\sum_{i=1}^{10}w^2_i+ \lambda^2- m_{11}^2 
\end{equation}
which is derived from the general inner product on the IIA roots
\begin{equation}
<\alpha,\beta>=\sum_{i=1}^{10}w^\alpha_i w^\beta_i+ \lambda^\alpha\lambda^\beta- m_{11}^\alpha m_{11}^\beta \label{IIAinnerproduct}
\end{equation}
where $\alpha=\sum_i w^{\alpha}_ie_i$ and $\beta=\sum_i w^{\alpha}_ie_i$ appearing at levels $(m_{11}^\alpha,\lambda^\alpha)$ and $(m_{11}^\beta,\lambda^\beta)$ respectively.
The problem of constructing the Young tableaux of the algebra $\mathfrak{e}_{11}$  becomes one of combining the basic two non-trivial tableaux: $\tiny\yng(1)$ (with $m_{11}=0, \lambda=-1$) and $\tiny\yng(1,1)$ ($m_{11}=1,\lambda=1$), to find generalised Young tableaux with $\beta^2=2,0,-2,\ldots$. The constituent $m_{11}$ and $\lambda$ values are added to find the level of the general tableau: $(m^\alpha_{11},\lambda^\alpha)+(m^\beta_{11},\lambda^\beta)=(m^\alpha_{11}+m^\beta_{11},\lambda^\alpha+\lambda^\beta)$. The procedure for constructing $\mathfrak{e}_{11}$ at level $(m_{11},\lambda)$ involves constructing all Young tableaux with $3m_{11}-\lambda$ boxes such that the widths of the Young tableaux, $w_i$ in (\ref{IIAinnerproduct}), satisfy $\beta^2=2,0,-2,\ldots$. For example at level $m_{11}=0$, in addition to the generators of $\mathfrak{sl}(10,\mathbb{R})$ we find by considering all possible Young tableaux shapes with $\lambda$ ranging from $-1$ to $m_{11}=0$, the tableaux associated to the D1 brane and the dilaton, as indicated in the first row of table \ref{table:IIA}. We can repeat this procedure level by level to quickly construct $\mathfrak{e}_{11}$, modulo the information about the multiplicity of the generators, as Young tableaux. For levels $0\leq m_{11} \leq 4$ the resulting Young tableaux are shown in table \ref{table:IIA}. For a fixed number of boxes in a Young tableau (i.e. at a fixed level in the decomposition) the calculation the root length squared may be shortened tremendously by noting that the movement of any box to a column to its immediate right raises the root length squared by two and the reverse movement of a box by one place to the left reduces the root length squared by two. 
\end{subsection}
\begin{subsection}{Type IIB string theory}
\begin{figure}[h]
\centering
\includegraphics[scale=0.8,angle=0]{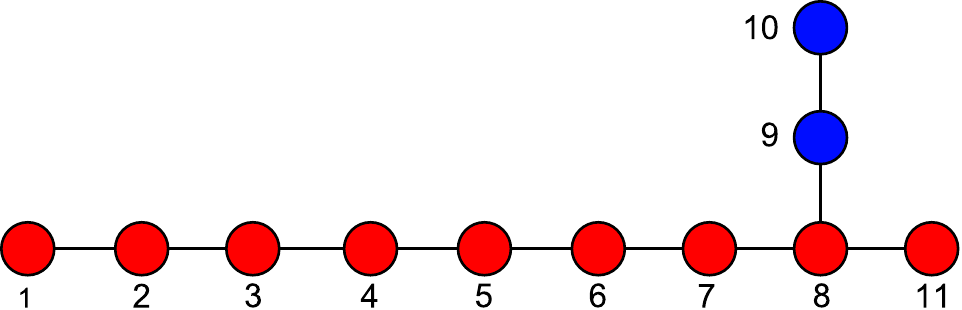}
\caption{The reduction of the Dynkin diagram for $E_{11}$} relevant to type-IIB theories  \label{E11IIB}
\end{figure}
The decomposition of $\mathfrak{e}_{11}$ into an algebra relevant to the type-IIB theory is found by deleting node nine from the Dynkin diagram of $E_{11}$ in figure \ref{E11IIB}. This leaves two disconnected Dynkin diagrams those of $SL(2,\mathbb{R})$, indicated by node 10, and $SL(10,\mathbb{R})$ consisting of nodes one to eight and node eleven, as labelled in figure \ref{E11IIB}. The generators of $\mathfrak{e}_{11}$ may now be written as a pair of Young tableaux, one corresponding to $SL(10,\mathbb{R})$ tensors and a second corresponding to $SL(2,\mathbb{R})$ tensors, and will include the generators of IIB supergravity at low levels. 

The decomposition strategy is slightly more involved than the case for the IIA theories due to the $SL(2,\mathbb{R})$ representations. For reference we give the Cartan sub-algebra of the decomposed algebra here:
\begin{align}
\nonumber H_i&={K^i}_i-{K^{i+1}}_{i+1} \qquad \mbox{for } i=1,2,\ldots 8,\\
H_9&=-\frac{1}{4}({K^1}_1+\ldots +{K^8}_8)+\frac{3}{4}({K^9}_9+{K^{10}}_{10})+\hat{R},\\
\nonumber H_{10}&=-2\hat{R} \qquad \mbox{and}\\
\nonumber H_{11}&={K^9}_9-{K^{10}}_{10}
\end{align}
where $\hat{R}$ is the generator associated to the dilaton in the IIB theory.
\begin{table}[ht] 
\centering 
\scalebox {0.95} {
\begin{tabular}{| c | c |} 
\hline Level, $m_{9}$ &  \\ [0.5ex]	
 \hline  
0 & $\bullet$ \\
& Dilaton \\
\hline &\\
1 & $(\,\Yvcentermath1\tiny\yng(1,1)\, ,\Yvcentermath1\tiny\yng(1)\,)$ \\
& D1 and F1 \\
\hline &\\
2 & $\left(\,\Yvcentermath1\tiny\yng(1,1,1,1)\, ,\Yvcentermath1\tiny\yng(1,1)\,\right)$ \\
 &D3\\
 \hline & \\
 3&$\left(\,\Yvcentermath1\tiny\yng(1,1,1,1,1,1)\, ,\Yvcentermath1\tiny\yng(2,1)\,\right)$\\ 
 &D5 and NS5\\ 
 \hline & \\
 4& $\left(\,\Yvcentermath1\tiny\yng(1,1,1,1,1,1,1,1)\, ,\Yvcentermath1\tiny\yng(2,2)\,\right)$\, , $\left(\,\Yvcentermath1\tiny\yng(1,1,1,1,1,1,1,1)\, ,\Yvcentermath1\tiny\yng(3,1)\,\right)$\, , $\left(\,\Yvcentermath1\tiny\yng(2,1,1,1,1,1,1)\, ,\Yvcentermath1\tiny\yng(2,2)\,\right)$\\
 & $\qquad  \qquad$ D7$_a$, NS7, D7$_b$ $\hspace{15pt} $ KK5\\
 \hline 
 \end{tabular} }
 \caption{Low level IIB Young tableaux from $\mathfrak{e}_{11}$. By $\bullet$ we indicate a Young tableau for a scalar. In addition the generators of $SL(10,\mathbb{R})$ and $SL(2,\mathbb{R})$ which are not indicated appear at level 0.} \label{table:IIB} 
\end{table}Roots in the algebra will be labelled by the level $m_9$ together with a root in the $SL(10,\mathbb{R})$ root lattice and another root in the $SL(2,\mathbb{R})$ root lattice. We will indicate the basis vectors of the $SL(10,\mathbb{R})$ root space by $\{f_1, f_2, \ldots, f_{10}\}$ and those of the $SL(2,\mathbb{R})$ root space by $\{g_1,g_2\}$.  Using this notation the simple roots of $E_{11}$ become:
\begin{align}
\nonumber \alpha_i&=f_i-f_{i+1}  \qquad \mbox{for } i=1,2,\ldots 8,\\
\alpha_9 &= f_9+f_{10}+g_2,\\
\nonumber \alpha_{10}&=g_1-g_2\qquad \mbox{and}\\
\nonumber \alpha_{11}&=f_9-f_{10}.
\end{align}
A general root of $E_{11}$ may be written:
\begin{align}
 \beta&=w_1f_1+w_2f_2+\ldots + w_{10}f_{10}+x_1g_1+x_2g_2 \label{IIBroot} \\
\nonumber & \equiv m_1\alpha_1+m_2\alpha_2+\ldots +m_{10}\alpha_{10}+m_{11}\alpha_{11}.
\end{align}
Such a root will correspond to a pair of Young tableaux: one $SL(10,\mathbb{R})$ Young tableau with rows of width $w_i$ where $i\in\{1,2,\ldots 10\}$ and one $SL(2,\mathbb{R})$ Young tableaux with rows of width $x_a$ where $a\in\{1,2\}$. The inner product decomposes to give:
\begin{equation}
\beta^2=\sum_{i=1}^{10}{w_i}^2+\sum_{a=1}^{2}{x_a}^2- m_9^2 \label{IIBinnerproduct}
\end{equation}
Note that the level $m_9$ is also the number of boxes in the $SL(2,\mathbb{R})$ Young tableau and half the number of boxes in the $SL(10,\mathbb{R})$ tableau, i.e. 
\begin{equation}
\#_{SL(2,\mathbb{R})}=m_9\qquad \mbox{and} \qquad \#_{SL(10,\mathbb{R})}=2m_9.
\end{equation}
The prescription for finding the algebra at level $m_9$ is to consider all the possible pairs of Young tableaux formed from $SL(10,\mathbb{R})$ tableaux with $2m_9$ boxes and $SL(2,\mathbb{R})$ tableaux with $m_9$ boxes. The algebra $\mathfrak{e}_{11}$ at level $m_9$ is encoded in those tableaux which satisfy $x_2\geq x_1$ and $w_{10}\geq w_9 \ldots w_2 \geq w_1$ and, at the same time, for which the corresponding (lowest weight) root $\beta$ satisfies $\beta^2=2,0,-2,\ldots$. We note again that we have given no consideration to the multiplicities of the generators appearing in $\mathfrak{e}_{11}$ and the complete, but significantly lengthier, construction of $\mathfrak{e}_{11}$ would include this information. The resulting roots are shown in table \ref{table:IIB}. 

We conclude this section by highlighting the pair of seven branes and pair of nine branes associated to real roots which we will denote D7$_a$, D7$_b$, D9$_a$ and D9$_b$. The D7$_a$ and D9$_a$ are Dirichlet branes whose tensions scale as $g_s^{-1}$, while the D7$_b$ and D9$_b$ have tensions which scale as $g_s^{-3}$ and $g_s^{-4}$ respectively. These branes are distinguished by particular $SL(2,\mathbb{R})$ tables within the $\bf{3}$ of seven branes and $\bf{4}$ of nine branes, hence here we indicate the $SL(2,\mathbb{R})$ labels for each brane:
\begin{align}
\beta_{D7_a}&=\left(\,\Yvcentermath1\tiny\yng(1,1,1,1,1,1,1,1)\, ,
\ytableausetup{mathmode, boxsize=1.0em}
 \begin{ytableau}
\mbox{{\scriptsize 2}}&\mbox{{\scriptsize 2}}&\mbox{{\scriptsize 2}}\\
\mbox{{\scriptsize 1}}  \\
\end{ytableau} 
\,\right), \qquad \beta_{D7_b}=\left(\,\Yvcentermath1\tiny\yng(1,1,1,1,1,1,1,1)\, ,
\ytableausetup{mathmode, boxsize=1.0em}
 \begin{ytableau}
\mbox{{\scriptsize 2}}&\mbox{{\scriptsize 1}}&\mbox{{\scriptsize 1}}\\
\mbox{{\scriptsize 1}}  \\
\end{ytableau} 
\,\right), \label{7branes}\\
\qquad \beta_{D9_a}&=\left(\,\Yvcentermath1\tiny\yng(1,1,1,1,1,1,1,1,1,1)\, ,\ytableausetup{mathmode, boxsize=1.0em}
 \begin{ytableau}
\mbox{{\scriptsize 2}}&\mbox{{\scriptsize 2}}&\mbox{{\scriptsize 2}}&\mbox{{\scriptsize 2}}\\
\mbox{{\scriptsize 1}}  \\
\end{ytableau} \,\right) \quad \mbox{and}  \quad \beta_{D9_b}=\left(\,\Yvcentermath1\tiny\yng(1,1,1,1,1,1,1,1,1,1)\, ,\ytableausetup{mathmode, boxsize=1.0em}
 \begin{ytableau}
\mbox{{\scriptsize 2}}&\mbox{{\scriptsize 1}}&\mbox{{\scriptsize 1}}&\mbox{{\scriptsize 1}}\\
\mbox{{\scriptsize 1}}  \\
\end{ytableau} \,\right). \label{9branes}
\end{align}
We recall that the number of boxes labelled $1$ in the $SL(2,\mathbb{R})$ tables, which is equal to the coefficient $x_1$ in \ref{IIBroot}, indicates the scaling of the associated object's tension which goes as $g_s^{-x_1}$ \cite{Cook:2008p936}.
\end{subsection}
\end{section}

\begin{section}{Marginal Bound States of String Theory}
The marginal bound states of string theory are formed from pairs of branes for which the gravitational background of the first brane exerts no force on a second brane statically-embedded in the background \cite{Tseytlin:1996p1683}. The bound state has zero binding energy. The no-force condition concurs with the harmonic superposition rule of Tseytlin \cite{Tseytlin:1996p1469} for the combination of brane metrics to give the metric of a marginal bound state. The derivation of the no-force condition is setting dependent as the action of the probe brane depends upon its nature (e.g. whether it is a brane of M-theory or string theory). A general rule for understanding when marginal brane intersections occurred was given in \cite{Argurio:1997p1652}, where conditions for a marginal bound state to be formed were given on the number of coincident world-volume directions were between two branes. These conditions were shown to be identical, in the context of Kac-Moody algebra, to the vanishing of the inner product between the two roots representing the branes \cite{Englert:2004p2875}. 

It is instructive to observe the direct relation between the force exerted on a probe brane and the inner product of $E_{11}$. To do this we may follow the construction of \cite{Tseytlin:1996p1683} and consider the 11-dimensional probe brane construction. The probe brane world-volume action (consisting of a Nambu-Goto action and a Wess-Zumino term) is added to the gravitational background action in 11 dimensions using a static embedding. The resulting action is expanded in powers of derivatives and the term with no derivatives gives the effective static potential
\begin{equation}
V=\sqrt{-\mbox{det}G_{mn}}+\frac{1}{(p+1)!}\epsilon^{m_1\ldots m_{p+1}}A_{m_1\ldots m_{p+1}}
\end{equation}
where $G_{mn}$ is the background 11-dimensional metric along the world-volume of the probe brane and $A_{m_1\ldots m_{p+1}}$ are the components of the background gauge field aligned with the world-volume of the probe p-brane. The brane solutions of M-theory are encoded in a group element \cite{West:2004p1593}\footnote{In \cite{West:2004p1593} the left-invariant Maurer-Cartan form $g^{-1}dg$ is used to calculate the vielbein ${(e^h)_\mu}^a$, while in this paper we use the right-invariant Maurer-Cartan form $dg g^{-1}$ which is preferred in \cite{Houart:2010p2928}. The difference in conventions introduces a multitude of minus signs in particular the veilbein becomes ${(e^{-h})_\mu}^a$ when using the right-invariant Maurer-Cartan form and we have made the change $h\rightarrow -h$ in equation \ref{solutiongeneratinggroupelement}.}
\begin{equation}
g_\beta=\exp(\frac{1}{2}\ln N (H\cdot \beta))\exp((1-N)E_\beta) \label{solutiongeneratinggroupelement}
\end{equation}
where $N$ is a harmonic function in the directions transverse to the brane, $\beta$ is a real root of $E_{11}$ and $E_\beta$ its associated generator in the algebra. The first exponential in the product encodes the vielbein and the scalar product is
\begin{align}
H\cdot \beta &= \sum_{i=1}^{11}H_im_i\equiv \sum_{i=1}^{11}h_i {K^i}_i
\end{align}
where $\beta=\sum_{i=1}^{11}m_i\alpha_i=\sum_{i=1}^{11}w_ie_i$ is a root of $E_{11}$ and $e_i$ are as defined in equation (\ref{IIAbasis}). The $h_i$ have a simple expression in terms of the widths $w_i$ of the associated Young tableau
\begin{equation}
h_i=-\frac{1}{9}\sum_{j=1}^{11}w_j+w_i=-\frac{1}{3}L+w_i
\end{equation}  
where $L=\frac{1}{3}\sum_{i=1}^{11}w_i$ is the level the root appears at in the decomposition into representations of $SL(11,\mathbb{R})$. The diagonal components of the metric are
\begin{equation}
G_{ii}=\exp(-\ln (N)h_i)=N^{-h_i}=N^{(\frac{1}{3}L-w_i)}.
\end{equation}
A second probe brane in this geometry whose root in $E_{11}$ is $\gamma=\sum_{i=1}^{11}w^{(2)}_ie_i$ will experience an effective potential with gravitational background contributions along its world volume directions, i.e. if we probed the background  with a $p$-brane oriented along spacetime directions $x^1, x^2, \ldots x^{p+1}$ we would need to evaluate
\begin{align}
\nonumber -\mbox{det}G_{mn}&=-G_{11}G_{22}\ldots G_{(p+1)(p+1)}\\
&=\prod_{i=1}^{p+1}N^{(\frac{1}{3}L-w_i)}.
\end{align}
The directions longitudinal to the probe brane associated to $\gamma$ correspond to the non-zero widths $w^{(2)}_i$ in its Young tableau and consequently we compute
\begin{equation}
-\mbox{det}G_{mn}=\prod_{i=1}^{11}N^{(\frac{1}{3}L-w_i)w^{(2)}_i}=N^{-<\beta,\gamma>}
\end{equation}
where $<\beta,\gamma>$ is the $E_{11}$ root space inner product. The expression for the effective potential becomes,
\begin{equation}
V=N^{-\frac{<\beta,\gamma>}{2}}-\delta(\beta-\gamma)N^{-1}
\end{equation}
where $\delta$ is the Kronecker delta function. Consequently if we probe a brane background with an identical, parallel brane so that $\gamma=\beta$ where $\beta$ is a real root we have $V=0$ and hence it exerts no static force on the probe brane. While if the probe brane differs from the brane sourcing the background geometry then,
\begin{equation}
V=N^{-\frac{<\beta,\gamma>}{2}}
\end{equation}
and the no force condition is satisfied if $<\beta,\gamma>=0$ giving a constant potential. Similar arguments may be constructed directly in the IIA and IIB theories as the inner product on the $E_{11}$ root space remains unchanged in each decomposition.

\begin{subsection}{Common brane intersections: $<\beta_p,\beta_q> =0$}
The intersecting branes of string theory correspond to pairs of roots one for each brane in the intersection whose inner product is zero. The common string theory intersections are $1_{NS}||5_{NS}$, $5_{NS}\perp 5_{NS}(3)$, $p\perp q (n)$\footnote{Where $p||q$ with ${p\leq q}$ denotes a $p$-brane within the worldvolume of a $q$-brane and $p\perp q(n)$ indicates a $p$-brane and $q$-brane with $n$ spatial directions in common.} where $n=\frac{1}{2} (p+q) -2$, $p\perp 1_{NS} (0)$ and $5_{NS}\perp p (n)$ where $n=p-1$\cite{Tseytlin:1996p1683,Bergshoeff:1997p1476}. Let us illustrate how these intersections are linked to orthogonal roots in the root space. Consider the example of the intersection of a $p$-brane and a $q$-brane in the IIA decomposition. The IIA root associated to a $p$-brane with longitudinal spacetime directions $x^{10-p},x^{11-p},\ldots x^{10}$ is
\begin{equation}
\beta_p^{IIA}=e_{10-p}+e_{11-p}+\ldots +e_{10} + (\frac{p}{2}-1)e_{11}. \label{IIApbraneroot}
\end{equation}
Singling out the coordinate $x^{10}$ to be timelike, then a $q$-brane having $n$ spatial directions in common with the p-brane is represented by the root
\begin{equation}
\beta^{IIA}_q=e_{10-p-q+n}+ \ldots +e_{9-p}+e_{10-n}+ \ldots +e_9+e_{10} + (\frac{q}{2}-1)e_{11}.
\end{equation}
Now, using (\ref{IIAinnerproduct}), 
\begin{align}
\nonumber <\beta^{IIA}_p,\beta^{IIA}_q> &=n+1+(\frac{p}{2}-1)(\frac{q}{2}-1)-\frac{pq}{4}\\
& =n+2-\frac{q+p}{2}
\end{align}
and hence their inner product is zero when $n=\frac{q+p}{2}-2$.

The highest weight root associated to a $Dp$-brane in the IIB theory takes the form
\begin{equation}
\beta^{IIB}_p=f_{10-p}+f_{10-p+1}+\ldots +f_{10} + g_1 + \frac{p-1}{2}g_2.
\end{equation}
A  $Dq$-brane root with $n$ spacetime directions in common with the worldvolume of the $p$-brane is 
\begin{equation}
\beta^{IIB}_q=f_{10-p-q+n}+\ldots+f_{9-q}+f_{10-n}+f_{10-n+1}+\ldots +f_{10} + g_1 + \frac{q-1}{2}g_2.
\end{equation}
Using the IIB inner product (\ref{IIBinnerproduct}) we again find
\begin{align}
\nonumber <\beta^{IIB}_p,\beta^{IIB}_q>&=n+2+\frac{1}{4}(p-1)(q-1)-\frac{1}{4}(p+1)(q+1)\\
& = n+2-\frac{p+q}{2}.
\end{align}
The reader may confirm that the inner product associated to the other intersections listed above are also zero using equations (\ref{IIAinnerproduct}) and (\ref{IIBinnerproduct}) on the roots depicted as Young tableaux in tables \ref{table:IIA} and \ref{table:IIB}.

We conclude by indicating the intersection rules for the D7$_b$ and D9$_b$ branes of type IIB theory, whose Young tableaux are shown in (\ref{7branes}) and (\ref{9branes}). We may, by checking that the the IIB inner product (\ref{IIBinnerproduct}) vanishes in each instance, add to the list of standard brane intersections : $F1\perp D7_b(0)$, $D3||D7_b$, $NS5\perp D7_b(4)$ and $NS5||D9_b$.
\end{subsection}
\begin{subsection}{Exotic brane intersections: $<\beta_p, \gamma>=0$}
The list of intersections of $p$ and $q$ branes given in the previous paragraphs exhausts the set of low-level roots $\beta_p$ and $\beta_q$ such that $<\beta_p, \beta_q>=0$. However at higher levels in the decomposition we may identify roots $\gamma$, which have no known association to solutions of IIA and IIB string theory, but which are also orthogonal to roots associated to $p$-branes. These exotic roots will give rise to a background geometry which exerts no force on a statically embedded $p$-brane. Later on in this paper we will interpret such higher level objects as bound states of branes oriented such that the constituent branes each form intersections which exert no force on a probe $p$-brane. In this section we find rules that identify which exotic brane intersections with the canonical branes exist. These constraints are expected to be useful for the construction of extremal black holes using exotic branes to build upon the canonical branes.
 
Let us denote a generic root $\gamma$ appearing at level $(m_\gamma,\lambda_\gamma)$ in the type-IIA decomposition by
\begin{equation}
\gamma =w_1e_1+w_2e_2+\ldots +w_{11}e_{11}.
\end{equation}
Using (\ref{IIAinnerproduct}) and (\ref{IIApbraneroot}) we find
\begin{equation}
<\beta_p, \gamma> = w_{10-p}+w_{11-p}+\ldots +w_{10}+w_{11}(\frac{p}{2}-1)-m_\gamma\frac{p}{2}.
\end{equation}
For the $D0$-brane we can find orthogonal roots $\gamma$ at level $m_\gamma$ if:
\begin{equation}
w_{10}=w_{11}.
\end{equation}
An example of such an exotic root occurs at level $m_\gamma=4$:
\begin{equation}
{\tiny\yng(2,2,1,1,1,1,1,1)} \label{IIA_eight_two}
\end{equation}
and is derived by dimensional reduction from the $[9,3]$ Young tableau in $D=11$, so it has $\lambda=w_{11}=2$. There are many more such real roots and hence in tables \ref{table:IIAexoticorthogonality} and  \ref{table:IIBexoticorthogonality} we indicate the conditions on the root $\gamma$ to be orthogonal to each of the canonical branes of the IIA and IIB theories.
\begin{table} 
\centering 
\begin{tabular}{| c | c |} 
\hline Brane. & Condition for orthogonality with $\gamma$. \\ [0.5ex]	
 \hline  
D0 & $w_{10}=w_{11}$  \\
\hline &\\
F1 & $m_\gamma=w_9+w_{10}+w_{11}$  \\
\hline &\\
D2& $m_\gamma=w_8+w_9+w_{10}$ \\
 \hline & \\
 D4 &  $m_\gamma=\frac{1}{2}(w_6+w_7+\ldots + w_{11})$ \\ 
 \hline & \\
 NS5 & $m_\gamma = \frac{1}{2}(w_5+w_6+\ldots + w_{10})$ \\
\hline & \\
 D6 & $m_\gamma=\frac{1}{3}(w_4+w_5+\ldots + w_{10}+2w_{11})$ \\
\hline & \\
 D8 &  $m_\gamma=\frac{1}{4}(w_2+w_3+\ldots + w_{10}+3w_{11})$ \\
 \hline 
 \end{tabular} 
 \caption{The condition for $\gamma\equiv\sum_{i=1}^{11}w_ie_i$ to be orthogonal to the root associated to the listed IIA branes. The exotic root $\gamma$ appears at level $(m_\gamma\equiv \frac{1}{3}(w_1+w_2+\ldots +w_{11}),\lambda\equiv w_{11})$ in the decomposition.}\label{table:IIAexoticorthogonality} 
\end{table}
\begin{table}
\centering 
\scalebox {0.95} {
\begin{tabular}{| c | c |} 
\hline Brane. & Condition for orthogonality with $\gamma$. \\ [0.5ex]	
 \hline  
D1 &  $x_1=w_9+w_{10}$ \\
\hline &\\
F1 &  $x_2=w_9+w_{10}+x_1$\\
\hline &\\
D3 &  $x_1+x_2=w_7+w_8+w_9+w_{10}$\\
 \hline & \\
D5 & $2x_1+x_2=w_5+w_6+\ldots+w_{10}$\\ 
 \hline & \\
 NS5 & $x_1+2x_2=w_5+w_6+\ldots+w_{10}$ \\
\hline & \\
 D7$_a$ & $3x_1+x_2=w_3+w_4+\ldots + w_{10}$\\
\hline & \\
 D9$_a$ & $2x_1=x_2$ \\
 \hline 
 \end{tabular} }
 \caption{The condition for $\gamma\equiv\sum_{i=1}^{10}w_if_i+x_1g_1+x_2g_2$ to be orthogonal to the root associated to the listed IIB branes.} \label{table:IIBexoticorthogonality} 
\end{table}
\end{subsection}
\end{section}
\begin{section}{Non-marginal bound states: $<\beta_p, \beta_q> \leq 0$.}
In this section we will focus on interpreting the exotic real roots as bound states of string theory branes. Suppose an exotic root $\gamma$ of $E_{11}$, may be decomposed as the sum of two brane roots $\gamma=\beta_p+\beta_q$. As $\gamma^2=\beta_p^2=\beta_q^2=2$ then $<\beta_p,\beta_q>=-1$. Consequently the static potential is not constant, the $p$-brane and $q$-brane exert a force on each other and the exotic root $\gamma$ is interpreted as a non-marginal bound state of a $p$-brane and a $q$-brane. All real roots appearing at level two or greater in the decomposition may be expressed as a sum of roots from lower levels and here we are interested in the case when the constituent roots are related to the standard branes of string theory so that
\begin{equation}
\gamma=\sum_{p,i}\beta_{p_i} 
\end{equation}
where $\gamma$ is a real root i.e. $\gamma^2=2$, $\beta_{p_i}$ is a root associated to a canonical string theory $p$ brane, the subscript $i$ distinguishes between $p$-branes with different orientations and we note that in most cases there are multiple ways to partition $\gamma$ into a sum of standard brane roots. For the IIA theory we take the canonical p-branes to be \{D0,F1,D2,D4,NS5,D6,D8\} branes, and for IIB we take the canonical branes to be \{D1,F1,D3,D5,NS5,D7$_a$,D7$_b$,D9$_a$,D9$_b$\} as labelled in section 2, so that $\beta_{D2_1}$ and $\beta_{D2_2}$ indicates the roots of two $D2$ branes with different orientations. The fact that $\gamma^2=2$ implies that the constituent roots $\beta_{p_i}$ have inner products satisfying $<\beta_{p_i},\beta_{q_j}>\leq 2$. In particular when $<\beta_{p_i},\beta_{q_j}>=-1$ or $0$ the roots $\beta_{p_i}$ form a set of simple positive roots of a simply-laced Dynkin diagram. Let us illustrate this by way of an example, consider the Young tableau in equation (\ref{IIA_eight_two}), it is associated to a real root $\gamma$ where 
\begin{equation}
\gamma = e_3+e_4+e_5+e_6+e_7+e_8+2e_9+2e_{10}+2e_{11}.
\end{equation}
and we may partition it in terms of roots of canonical branes in a number of different ways, for example, we find by inspection that:
\begin{equation}
\gamma=\beta_{D6}+\beta_{D2}
\end{equation}
where
\begin{align}
\beta_{D6}&=e_4+e_5+e_6+e_7+e_8+e_9+e_{10}+2e_{11}, \quad \mbox{and} \\
\nonumber \beta_{D2}&=e_3+e_9+e_{10}.
\end{align}
We may confirm that $\beta_{D6}\cdot \beta_{D2}=-1$ and the truncation of the algebra $\mathfrak{e}_{11}$ to the roots $\beta_{D2}$,  $\beta_{D6}$ and $\beta_{D2}+\beta_{D6}=\gamma$ gives positive root system identified with that of an $SL(3,\mathbb{R})$ embedded in $E_{11}$. We may quickly identify other sums of canonical brane roots that also give the exotic brane root $\gamma$, for example a partition with three brane roots is
\begin{equation}
\gamma=\beta_{D4}+\beta_{D2}+\beta_{F1}
\end{equation}
where
\begin{align}
\beta_{D4}&=e_6+e_7+e_8+e_9+e_{10}+e_{11}, \\
\nonumber \beta_{D2}&=e_3+e_9+e_{10}  \quad \mbox{and}\\
\nonumber \beta_{F1}&=e_4+e_5+e_{11}.
\end{align}
Now we note that $\beta_{D4}\cdot \beta_{D2}=0$, $\beta_{D4}\cdot \beta_{F1}=-1$, $\beta_{D2}\cdot \beta_{F1}=-1$ and hence the roots $\beta_{F1}$, $\beta_{D2}$, $\beta_{D4}$, $\beta_{D4}+\beta_{F1}$, $\beta_{F1}+\beta_{D2}$ and $\beta_{D4}+\beta_{F1}+\beta_{D2}=\gamma$ are the positive roots of $SL(4,\mathbb{R})$ embedded in $E_{11}$. We may continue and find other ways to partition $\gamma$ into brane roots. To summarise we may identify the root $\gamma$ with a single exotic brane solution associated to an $SL(2,\mathbb{R})$ embedded in $E_{11}$ or to a composite object comprised of an interacting D2 and D6 brane which associated to an embedding of $SL(3,\mathbb{R})$ in $E_{11}$ or to a composite object formed from a D4, D2 and F1 brane associated to an embedding of $SL(4,\mathbb{R})$ in $E_{11}$ as well as other more complicated descriptions. The point of view adopted in this paper is that each of these descriptions will describe in some limit the same fundamental exotic object and in this way we interpret mixed symmetry tensors as bound states of standard branes. We may also see from the example that it is a non-trivial problem to find all partitions of an exotic root into standard brane roots by hand.
 
Of particular interest are the cases where the Dynkin diagram formed by the canonical brane roots corresponds to that of a semisimple Lie group ${\cal G}$. In this case the gravitational background corresponding to the exotic root $\gamma$ may be described using a one-dimensional $\sigma$-model on a coset $\frac{{\cal G}}{{\cal K(G)}}$ \cite{Houart:2010p2928} where ${\cal K(G)} < {\cal G}$.  Let us digress from discussing the $\sigma$-model to make some comments on ${\cal K(G)}$.  When ${\cal G}=E_{11}$ the sub-group ${\cal{K}}(E_{11})$ is determined from the choice of local space-time signature that defines a temporal involution: it is the sub-group of ${\cal G}$ whose algebra is invariant under the temporal involution. We recall that the Chevalley-Cartan involution leaves invariant the maximal compact sub-algebra of a finite algebra while the temporal involution leaves invariant the algebra of a non-compact sub-group. For example one may single out an $\mathfrak{so}(1,9)$ sub-algebra within $\mathfrak{sl}(10,\mathbb{R})$ using the space-time signature, this defines the action of a temporal involution on the generators associated to the simple roots of $SL(10,\mathbb{R})$ by $\Omega$ as $\Omega (E_i)=-\eta_{ij}F_j$ where $E_i$ denotes a generator associated to a positive simple root $\alpha_i$, $F_j$ a generator associated to a negative simple root $-\alpha_j$ and $\eta_{ij}$ is the Minkowski metric written using the convention that timelike vectors have negative length squared. This is sufficient to define the temporal involution on all the generators associated to positive roots as the temporal involution distributes over the commutator: $\Omega [E_i,E_j]=[\Omega(E_i),\Omega(E_j)]\equiv \Omega(E_{(i+j)})$. In the same manner the temporal involution defined on the generators of the positive, simple roots of $E_{11}$ is also defined on all the generators associated to positive roots of $E_{11}$. For the IIA algebra we define $\Omega(R^a)=-(-1)^nR_a$, $\Omega(R^{a_1a_2})=-(-1)^nR_{a_1a_2}$, $\Omega(R)=-R$ and $\Omega(H_i)=-H_i$ where $n$ is the number of temporal indices on the generator\footnote{E.g. Let the tenth coordinate index label the single time-like direction in a background with signature $(1,9)$ then $\Omega(R^{9 10})=R_{910}$ while $\Omega(R^{89 10})=-R_{89}$}, while for the IIB algebra we define $\Omega(R^{a_1a_2})=-(-1)^nR_{a_1a_2}$, $\Omega({K^{\bar{1}}}_{\bar{2}})=-{K^{\bar{2}}}_{\bar{1}}$, $\Omega(\hat{R})=-\hat{R}$ and $\Omega(H_i)=-H_i$, where ${K^{\bar{1}}}_{\bar{2}}$ is the generator associated to the axion and again $n$ is the number of temporal indices on the generator.

The global ${\cal G}$ of the $\sigma$-model is generated by the relevant orientation of the branes in the bound state, and the action of ${\cal K(G)}$ on the bound solution is to mix the relative charges of the constituent branes. Earlier we discussed the example of an exotic root $\gamma$ which could be expressed as a sum of canonical brane roots $\beta_{p_i}$, but we may also decompose a canonical $p$-brane root $\beta_{p_i}$ into a sum of roots associated to canonical $p'$-branes where $p'<p$. The decomposition of canonical $p$-brane roots will be emphasised in this paper - as it will make contact with the string theory literature and will develop the case for a similar treatment of the exotic roots. Take, by way of a simple brane example, the dyonic membrane in M-theory \cite{Izquierdo:1995p1636} which is generated from two roots, $\beta_{M2}$ and $\beta_{S2}$ associated to an $M2$ brane and its Euclidean counterpart the $S2$ brane \cite{Cook:2009p2751,Houart:2010p2928}. The roots satisfy the condition that $\beta_{M2}\cdot \beta_{S2}=-1$ and that $\beta_{M2}+\beta_{S2}=\beta_{M5}$. The negative inner product is enough to ensure that the commutator of the corresponding generators does not terminate, by the Serre relations, and that, up to multiplicity, the sum of the roots, here $\beta_{M5}$, is also a root in the root system of $E_{11}$. The dyonic membrane possesses a manifest $SL(3,\mathbb{R})$ symmetry in this setting \cite{Houart:2010p2928}. The symmetry can be understood by taking the roots involved in the bound states as simple positive roots of a Lie algebra $\mathfrak{g}$ of the group ${\cal G}$. The Cartan involution for the M-theory generators in this example is $\Omega (E_{M2})=F_{M2}$, $\Omega (E_{S2})=-F_{S2}$ and $\Omega (E_{M5})=-F_{M5}$ - these are derived from the action of the involution $\Omega$ on the generators associated to the simple roots of $E_{11}$. The normalised invariant sub-algebra is $\frac{1}{2}(E_{M2}+F_{M2})$, $\frac{1}{2}(E_{S2}-F_{S2})$ and $\frac{1}{2}(E_{M5}+F_{M5})$ which is the algebra $\mathfrak{so}(2,1)$. The action of the subgroup in the coset action ${\cal K(G)} = SO(2,1)$ changes the contribution of $M2$ and $M5$ charge in the bound state. Starting with an $M2$ brane and acting with an element of ${\cal K(G)}$ gives the dyonic membrane with a particular interpolating angle.

In this section we outline the procedure for finding all possible bound states of the canonical branes present in both the IIA and IIB string theories. The results will be too numerous to present here in their entirety but we will give a summary of the findings and discuss the most interesting examples. The complete list of our results will be uploaded to the arxiv with the preprint of the present article. The analysis of the results is split into three sections. In the sections 4.1 and 4.2 we discuss the low rank groups embedded in $E_{11}$ up to rank 4 for IIA and IIB string theory respectively. Section 4.3 consists of a general discussion of cosets on groups of rank 5 and above which is relevant to both types IIA and IIB. 

\begin{subsection}{Low rank IIA string theory bound states.}
Our aim will be to find all the recognisable Cartan matrices $A_{ij}=<\beta_i,\beta_j>$ where $\beta_i$  are roots of $E_{11}$ associated to IIA D-branes, NS-branes and S-branes. Specifically we will consider bound states formed of the following branes and their Euclidean (S-brane) counterparts: the D0, F1, D2, D4, NS5, D6 and D8 branes. The derivation of the roots associated to these branes is given in section 1.1 and their Young tableaux, from which the usual root expansion may be read, is shown in table \ref{table:IIA}. The process of finding the Cartan matrices begins by taking the seven canonical branes to each be associated to an embedding of $SL(2,\mathbb{R})$ in $E_{11}$, which we note has a rank one Cartan matrix. By then systematically searching through all the different canonical brane orientations, to find if one may be added to the single brane such that a recognisable rank two Cartan matrix is constructed. Once all the rank $n$ symmetries of bound states have been found the process may be continued to find the rank $n+1$ symmetries. A small and finite set of the Cartan matrices constructed in this way will correspond to simply-laced Dynkin diagrams which may be treated using the coset formalism detailed in \cite{Houart:2010p2928}. We will not discuss in detail any solutions related to affine Kac-Moody algebras in what follows. For the set of canonical states used in this construction the largest simple Dynkin diagram (containing no loops and no more than one root with three bonds) is found at rank ten. This, however, contains many exotic states, roots $\gamma$ whose association with the canonical branes is not understood. The largest Cartan matrix for which all positive roots are associated to canonical branes is occurs at rank four. Here we will outline some of the interesting cases up to and including rank four and subsequently we will discuss the larger symmetry bound states containing exotic content.
\end{subsection}

\begin{subsubsection}{Type-IIA: Cosets on rank two groups.}
There is only one finite rank two simply-laced Dynkin diagram, that associated to $SL(3,\mathbb{R})$:
\begin{equation}
\nonumber \includegraphics[scale=0.65,angle=0]{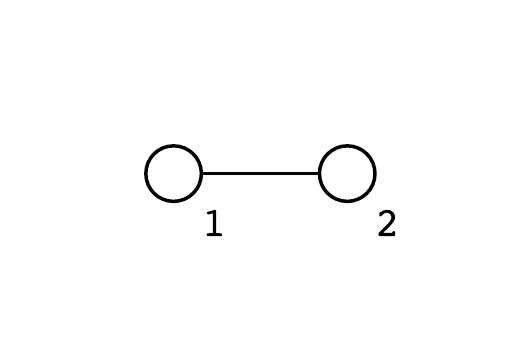}
\end{equation}
Within the roots of $E_{11}$ decomposed into the IIA representations there are forty-five bound states possessing an $SL(3,\mathbb{R})$ symmetry. Eleven of these involve only roots associated to standard type-IIA  solutions and we refer to such bound states as pure bound states. These pure states are well known, although their interpretation as an $SL(3,\mathbb{R})$ bound state may not be, and we list them for reference in table \ref{IIA_rank_2_bound_states}. 
\begin{table}[ht]
\centering
\begin{tabular}{ c || c | c | c | c | c | c | c }
& S0 & S1 & S2 & S4 & S5 & S6 & S8 \\
\hline
\hline
D0& - & (D0,D2) & - & (D0,NS5) & - & - & -  \\
\hline
F1 & (F1,D2) & - & (F1,D4) & (F1,D6) & - & (F1,D8) & -  \\
\hline
D2 & - & (D2,D4) & (D2,NS5) & - & - & - & -  \\
\hline
D4 & (D4,NS5) & (D4,D6) & - & - & - & - & -  \\
\hline
NS5 & - & - & - & - & - & - & -  \\
\hline
D6 & - & (D6,D8) & - & - & - & - & -  \\
\hline
D8 & - & - & - & - & - & - & -  \\
\end{tabular}
\caption{The $SL(3,\mathbb{R})$ bound states of the canonical IIA branes. Along the first column and the first row we list the branes acting as simple roots of the $SL(3,\mathbb{R})$ symmetry. At the intersection of row and column we give the second brane in the bound state where such a state exists. E.g. a bound state of a D0 and a D2 brane is found by combining the simple roots for a D0 brane and an S1 brane with no common worldvolume directions.} \label{IIA_rank_2_bound_states}
\end{table}
The remaining bound states with an $SL(3,\mathbb{R})$ symmetry include an exotic root, one to which no type-IIA  solution is associated, however they may be treated using the same techniques that lead to the pure bound states solutions. Bound states of pure and exotic types of branes are all described by a null geodesic path on the coset space $\frac{SL(3,\mathbb{R})}{SO(1,2)}$ and the generic solution in terms of harmonic function for each rank two case is given in \cite{Houart:2010p2928} where the (D6,D8) bound state is investigated in detail. By way of example of the general method we consider the (F1,D2) pure bound state formed from roots
\begin{equation}
\beta_{S0}=e_8-e_{11} \quad \mbox{ and } \quad \beta_{F1}=e_9+e_{10}+e_{11}.
\end{equation}
Note that $\beta_{D2}=e_8+e_9+e_{10}=\beta_{F1}+\beta_{S0}$. The coset model has the Lagrangian  
\begin{equation}
{\cal L}= (P_\xi|P_\xi)
\end{equation}
where $(M|N)\equiv Tr (MN)$ is the Killing form on $E_{11}$ and $P_\xi$ is the part of the Maurer-Cartan one-form component $\nu_\xi$ complementary to the Borel sub-algebra of $\mathfrak{so}(1,2)$, denoted $Q_\xi$, i.e.
\begin{equation}
\nu_\xi =(\partial_\xi g) g^{-1}=P_\xi+Q_\xi.
\end{equation}
The equation of motion for the Lagrangian, $\cal L$, is
\begin{equation}
\partial_\xi P_\xi - [Q_\xi,P_\xi]=0. \label{eom}
\end{equation}
A representative coset element $g$ is given by
\begin{equation}
g=\exp(\phi_1(\xi)H_1+\phi_2(\xi)H_2)\exp(C_1(\xi)E_1+C_2(\xi)E_2+C_{12}(\xi)E_{12}) \label{sl2solution1}
\end{equation}
and the generators which form the Borel sub-algebra of $\mathfrak{sl}(3,\mathbb{R})$ are denoted
\begin{align}
\nonumber H_1&=-\frac{1}{8}({K^1}_1+\ldots +{K^7}_7+{K^9}_9+{K^{10}}_{10})+\frac{7}{8}{K^8}_8-\frac{3}{2}{R} ,\\
H_2&=-\frac{1}{4}({K^1}_1+\ldots + {K^8}_8)+\frac{3}{4}({K^9}_9+{K^{10}}_{10})+{R},\\
\nonumber E_1&=R^{8}, \quad E_2=R^{910} \quad \mbox{and} \quad E_{12}=R^{8910}.
\end{align}
The temporal involution, $\Omega$, defined on $\mathfrak{e}_{11}$ to pick out a temporal coordinate in the background spacetime has a pre-defined action on the generators of the embedded $\mathfrak{sl}(3,\mathbb{R})$. If we pick the temporal coordinate to be $x^{10}$ we find that 
\begin{equation}
\Omega{(E_1)}=-F_1, \quad \mbox{and} \quad  \Omega{(E_2)}=F_2.
\end{equation}
The local group is the real form of $SO(3)$ left invariant by $\Omega$ and is therefore $SO(1,2)$. The equation of motion (\ref{eom}) of the Lagrangian together with the quadratic Hamiltonian constraint $(P_\xi | P_\xi)=0$ gives the condition for a null geodesic motion on the coset  $\frac{SL(3,\mathbb{R})}{SO(1,2)}$ and the solution \cite{Houart:2010p2928} is given in terms of two harmonic functions in the real parameter $\xi$, $N_1$ and $N_2$ 
\begin{align}
\nonumber \phi_1&=\frac{1}{2}\ln(N_1), \quad \phi_2=\frac{1}{2}\ln(N_2),\\
C_1&=\frac{\tan{\beta}}{N_1}, \quad C_2=\frac{\sin{\beta}}{N_2}\quad \mbox{and} \quad C_{12}=\frac{1}{2\cos{\beta}}(\frac{\cos^2{\beta}}{N_1}+\frac{1}{N_2}) \label{sl2solution2}
\end{align}
where $N_1(\xi)=1+q\xi\cos^2{\beta}$ and $N_2(\xi)=1+q\xi$ and $\beta\in [0,\frac{\pi}{2}]$. The parameter $\beta$ is associated to the action of the compact generator in $SO(1,2)$ and controls the charge of the F1 string; as $\beta$ varies the solution interpolates between the F1 string and a D2 brane. The metric, deduced from $\phi_1$ and $\phi_2$, in the Einstein frame is
\begin{equation}
ds^2=N_1^{\frac{1}{8}}N_2^{\frac{1}{4}}((dx^1)^2+\ldots+(dx^7)^2+N_1^{-1}(dx^8)^2+N_2^{-1}(-(dt^9)^2+(dx^{10})^2)). 
\end{equation}
When $\beta\rightarrow \frac{\pi}{2}$ $N_1\rightarrow 1$ and the metric reduces to that of an F1 string while when $\beta\rightarrow 0$ $N_1\rightarrow N_2$ and the metric becomes that of a D2 brane. The dilaton is found from the coefficient of $R$ in the group element
\begin{equation}
e^A=\exp(\frac{3}{4}\ln(N_1)-\frac{1}{2}\ln(N_2))=N_1^{\frac{3}{4}}N_2^{-\frac{1}{2}}.
\end{equation}
The embedding of the coset in spacetime is achieved by identifying the parameter $\xi$ with one of the space-time directions $x^i$ transverse to the S0, F1 and D2 branes. As the functions are harmonic in a single transverse coordinate the solution is a one-dimensional smeared solution. One may localise the solution by unsmearing the harmonic functions, in the directions transverse to all constituent branes, so that $N_1$ and $N_2$ become 
\begin{align}
N_1=1+\frac{q\cos^2{\beta}}{r^5} \quad \mbox{and} \quad N_2=1+\frac{q}{r^5}
\end{align}
where $r^2=(x^1)^2+(x^2)^2+\ldots+(x^7)^2$.
The active field strength components are\footnote{In the notation of \cite{Houart:2010p2928} the string theory field strengths are $F_{\xi,i}=(e^A)^{a_i}P_{\xi,i}$ where $e^A$ is the dilaton and $[R,E_i]=a_iE_i$. For this example we use $a_1=-\frac{3}{4}$, $a_2=\frac{1}{2}$ and $a_{12}=-\frac{1}{4}$ which are deduced from $[H_i,E_i]=2E_i$.}
\begin{align}
\nonumber G_{i 8}&=\tan{\beta}\,\partial_i N_1^{-1}, \\
H_{i 910}&=\sin{\beta}\, \partial_i N_2^{-1} \qquad \mbox{and} \\
\nonumber F_{i 8910}&=-\cos{\beta}\frac{\partial_iN_2}{N_1N_2}
\end{align}
where $i$ indicates a coordinate transverse to all the component branes, i.e. $i\in \{1,2,\ldots 7\}$. All the solutions in table \ref{IIA_rank_2_bound_states} can be constructed in a similar way using the results of \cite{Houart:2010p2928}. This example is reminiscent of the supertube \cite{Mateos:2001p3100} as the parameter $\beta$ controls the "turning on" of a D0 brane charge in the background of a fundamental string to blow it up into a D2 brane.
\end{subsubsection}
\begin{subsubsection}{Type-IIA: Cosets on rank three groups.}
There are two simply-laced Dynkin diagrams of rank three, corresponding to $SL(4,\mathbb{R})$ and the affine algebra $SL(3,\mathbb{R})^+$. Their respective Dynkin diagrams are:
$$\includegraphics[scale=0.5,angle=0]{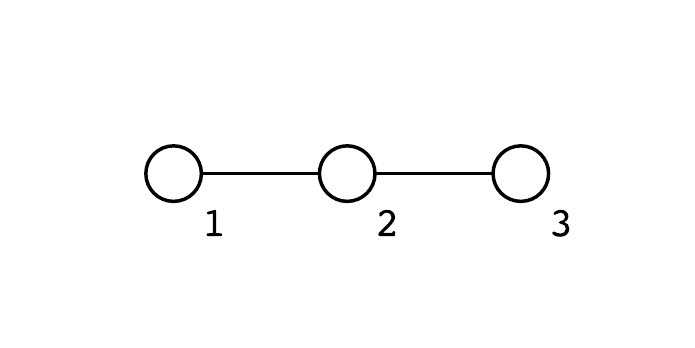}\qquad \qquad \includegraphics[scale=0.5,angle=0]{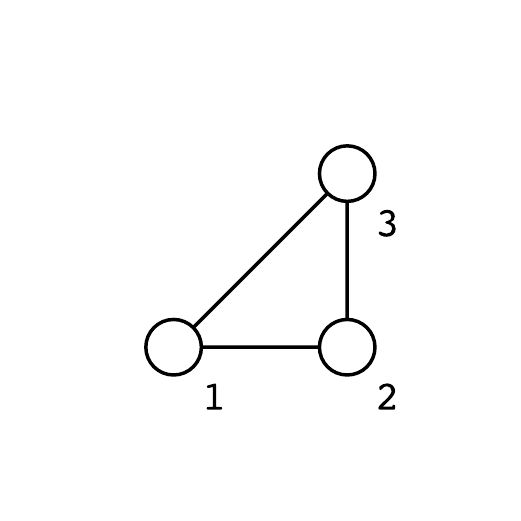}$$
There are 294 type-IIA  bound states which possess an $SL(4,\mathbb{R})$ symmetry, of these only nine bound states may be interpreted as pure bound states of branes, these cases are indicated in table \ref{IIA_rank_3_bound_states_finite}. In addition there are eleven pure bound states corresponding to the embedding of $SL(3,\mathbb{R})^+$ in $E_{11}$ listed in table  \ref{IIA_rank_3_bound_states_affine} we will not discuss these examples further here.
\begin{table}[ht]
\centering
\begin{tabular}{ c || c | c | c  }
& S0 & S1 & S2   \\
\hline
\hline
(D0,D2) & - & (D0,D2,D2,D4) & (D0,D2,NS5)  \\
\hline
(F1,D2)  & - & (F1,D2,D4) & (F1,D2,D4,NS5)   \\
\hline
(F1,D4) & - & (F1,D4,D6) & -   \\
\hline
(F1,D6) & - & (F1,D6,D8) & - \\
\hline
(D2,D4) & (D2,D4,NS5) & (D2,D4,D4,D6) & -  \\
\hline
(D4,D6) & - & (D4,D6,D6,D8) & -   \\
\end{tabular}
\caption{The $SL(4,\mathbb{R})$ bound states of the canonical IIA branes. Adding an S-brane indicated by the first row to an $SL(3,\mathbb{R})$ bound state in the first column gives a bound state comprised of the branes indicated at the intersection. Other combinations give rise to bound states of a different symmetry or bound states including an exotic brane.} \label{IIA_rank_3_bound_states_finite}
\end{table}
\begin{table}[ht]
\centering
\begin{tabular}{ c || c | c | c | c  }
& S1 & S2 & S4 & S6 \\
\hline
\hline
(D0,D2) & - & - & (D0,D2,NS5) & - \\
\hline
(D0,NS5) & - & - & (D0,NS5,NS5) & - \\
\hline
(F1,D2)  & - & - & (F1,D2,D6) & (F1,D2,D8) \\
\hline
(F1,D4) & - & (F1,D4,D4) & (F1,D4,D6) & (F1,D4,D8) \\
\hline
(F1,D6) & - & - & (F1,D6,D6) & - \\
\hline
(D2,D4) & - & (D2,D4,NS5) & - & - \\
\hline
(D2,NS5) & - & (D2,NS5,NS5) & - & - \\
\hline
(D4,NS5) & (D4,NS5,D6) & - & - & -  \\
\end{tabular}
\caption{The $SL(3,\mathbb{R})^+$ bound states of the canonical IIA branes. Adding an S-brane indicated by the first row to an $SL(3,\mathbb{R})$ bound state in the first column gives a bound state comprised of the branes indicated at the intersection.} \label{IIA_rank_3_bound_states_affine}
\end{table}
Of the $SL(4,\mathbb{R})$ bound states those which include the D8 brane have been discussed in detail in  \cite{Houart:2010p2928} and a solution for any coset of $SL(4,\mathbb{R})$ has been given there. By way of example we shall indicate how the (D0,D2,D2,D4) bound state solution may be found from the one-dimensional coset model. The three simple roots involved in the bound state are:
\begin{align}
\nonumber \beta_{S1(1)}&=e_8+e_9+e_{11}, \\
\beta_{S1(2)}&=e_{10}-e_{11} \qquad \mbox{and} \\
\nonumber \beta_{D0}&=e_6+e_7+e_{11}. 
\end{align}
The representative coset element is:
\begin{align}
\nonumber g= \exp(\phi_1(\xi)H_1+\phi_2(\xi)H_2+\phi_3(\xi)H_3)\exp(&C_1(\xi)E_1+C_2(\xi)E_2+C_3(\xi)E_3\\
\nonumber+&C_{12}(\xi)E_{12}+C_{23}(\xi)E_{23}\\
+&C_{123}(\xi)E_{123})
\end{align}
and the generators forming the Borel sub-algebra of $\mathfrak{sl}(4,\mathbb{R})$ are
\begin{align}
\nonumber H_1&=-\frac{1}{4}({K^1}_1+\ldots +{K^7}_7+{K^{10}}_{10})+\frac{3}{4}({K^8}_8+{K^9}_9)+R,\\
H_2&=-\frac{1}{8}({K^1}_1+\ldots + {K^9}_9)+\frac{7}{8}{K^{10}}_{10}-\frac{3}{2}{R},\\
\nonumber H_3&=-\frac{1}{4}({K^1}_1+\ldots +{K^5}_5+{K^8}_8+\ldots +{K^{10}}_{10})+\frac{3}{4}({K^6}_6+{K^7}_7)+R,\\
\nonumber E_1&=R^{89}, \quad E_2=R^{10}, \quad E_{3}=R^{67},\\
\nonumber E_{12}&=R^{8910}, \quad E_{23}=R^{6710} \quad \mbox{and}\\
\nonumber E_{123}&=R^{678910}.
\end{align}
By applying the commutators of $E_{11}$ we see that the generators close on themselves with
\begin{equation}
[E_1,E_2]=E_{12},\quad [E_2,E_3]=E_{23} \,\, \mbox{ and } \,\, [E_1,E_{23}]=[E_{12},E_{3}]=E_{123}
\end{equation}
which are the commutation relations of $\mathfrak{sl}(4,\mathbb{R})$. In our identification of the roots with electric and spacelike branes we have implicitly chosen the $x^{10}$ coordinate to be timelike. The corresponding temporal involution is $\Omega$ defined on the generators associated to the positive simple roots by
\begin{equation}
\Omega(E_1)=-F_1, \quad \Omega(E_2)=F_2, \quad \mbox{ and } \quad \Omega(E_3)=-F_3
\end{equation}
where $F_1$, $F_2$ and $F_3$ are the generators for the negative simple roots. This temporal involution leaves an $\mathfrak{so}(2,2)$ sub-algebra invariant, and the solution to the Lagrangian equation of motion in this rank four example is described by a null geodesic on an $\frac{SL(4,\mathbb{R})}{SO(2,2)}$ coset. The null geodesic motion on the coset, which solves the Lagrangian equation of motion, is given by
\begin{align}
\nonumber \phi_1&=\frac{1}{2}\ln N_1, \quad \phi_2=\frac{1}{2}\ln N_2, \quad \phi_3=\frac{1}{2}\ln N_3\\ 
C_1&=-\frac{\tan{\beta}}{N_1}, \quad C_2=\frac{\sin{\beta}\sin{\gamma}}{N_2}, \quad C_3=-\frac{\tan{\gamma}}{N_3}\\
\nonumber C_{12}&=\frac{\sin{\gamma}}{2\cos{\beta}}(\frac{\cos^2{\beta}}{N_1}+\frac{1}{N_2}), \quad C_{23}=-\frac{\sin{\beta}}{2\cos{\gamma}}(\frac{\cos^2{\gamma}}{N_3}+\frac{1}{N_2}) \mbox{ and}\\
\nonumber C_{123}&=-\frac{1}{3\cos{\gamma}\cos{\beta}}(\frac{\cos^2{\beta}}{N_1}+\frac{1}{2N_2}+\frac{\cos^2{\gamma}}{N_3}+\frac{N_2\cos^2{\beta}\cos^2{\gamma}}{2N_1N_3})
\end{align}
where $N_1=1+q\cos^2{\beta}\xi$, $N_2=1+q\xi$ and $N_3=1+q\cos^2{\gamma}\xi$ are three harmonic functions in one-dimension and $\beta,\gamma\in[0,\frac{\pi}{2}]$.  The metric for the solution is diagonal and can be read directly from $\phi_1$, $\phi_2$ and $\phi_3$ to be
\begin{align}
\nonumber ds^2=(N_1N_3)^{\frac{1}{4}}N_2^{\frac{1}{8}}(&(dx^1)^2+\ldots+(dx^5)^2+N_3^{-1}((dx^6)^2+(dx^7)^2)\\
&+N_1^{-1}((dx^8)^2+(dx^9)^2)+N_2^{-1}(dx^{10})^2).
\end{align}
The dilaton is found from the coefficient of $R$ in the group element
\begin{equation}
e^A=\frac{N_2^{\frac{3}{4}}}{(N_1N_3)^{\frac{1}{2}}}
\end{equation}
and we note that as $\beta$ and $\gamma$ vary $e^A$ ranges between $N_2^{\frac{3}{4}}$ (for $\beta=\gamma=\frac{\pi}{2}$) and $N_2^{-\frac{1}{4}}$ (for $\beta=\gamma=0$). The  coordinate $\xi$ may be identified with a space-time direction, here we choose $\xi=x^1$ in order to explicitly write the non-trivial field-strength components
\begin{align}
\nonumber G=&\partial N_2^{-1}\sin{\beta}\sin{\gamma}dx^1\wedge dx^{10}\\
\nonumber H=&-\partial N_1^{-1}\tan{\beta}dx^1\wedge dx^8\wedge dx^9-\partial N_3^{-1}\tan{\gamma}dx^1\wedge dx^6\wedge dx^7\\
F=&-\frac{\partial N_2}{N_1N_2}\cos{\beta}\sin{\gamma}dx^1\wedge dx^8\wedge dx^9\wedge dx^{10}\\
\nonumber &+\frac{\partial N_2}{N_2N_3}\sin{\beta}\cos{\gamma}dx^1\wedge dx^6\wedge dx^7\wedge dx^{10}\\
\nonumber &+e^{\frac{A}{2}}\partial N_2 \cos{\beta}\cos{\gamma} dx^2\wedge dx^3 \wedge dx^4 \wedge dx^5
\end{align}
\end{subsubsection}
\begin{subsubsection}{Type-IIA: Cosets on rank four groups.}
There are six inequivalent, simply-laced Dynkin diagrams of rank four:
\begin{align}
\nonumber &\includegraphics[scale=0.5,angle=0]{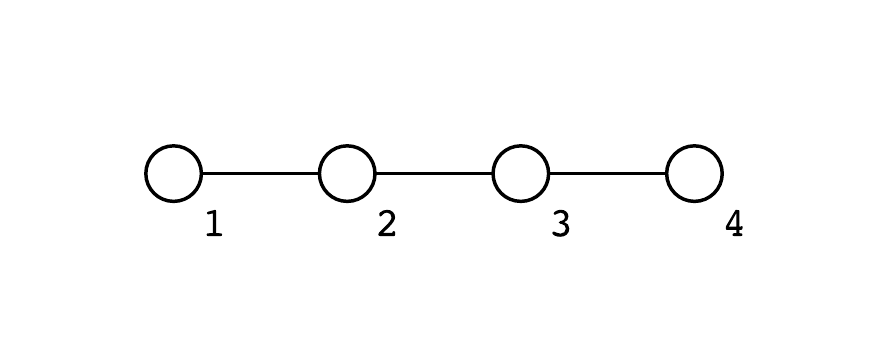} \includegraphics[scale=0.5,angle=0]{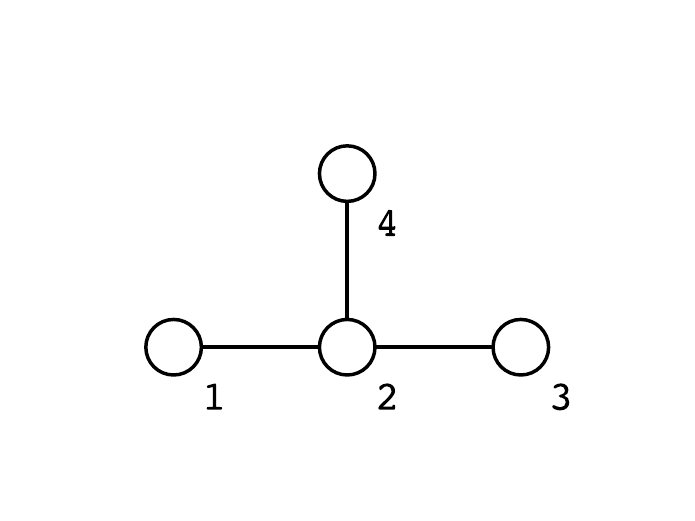}\includegraphics[scale=0.5,angle=0]{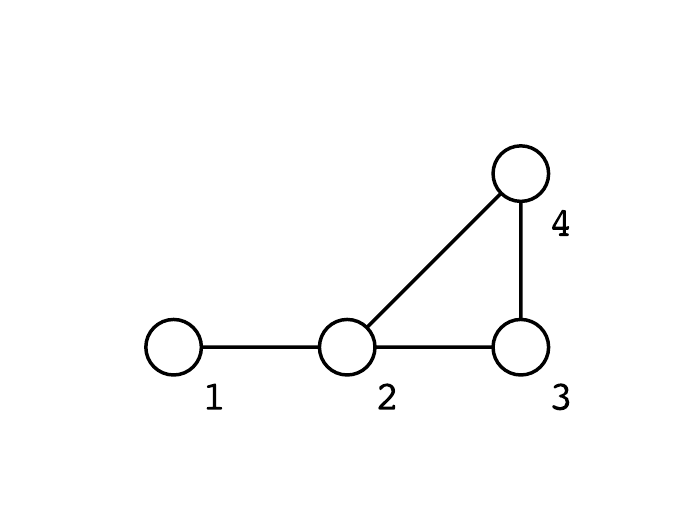}\\
 \nonumber &\,\, \quad \includegraphics[scale=0.5,angle=0]{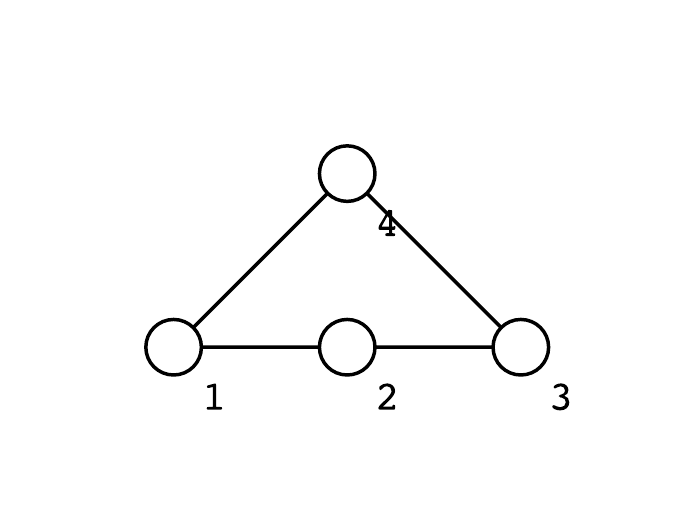} \quad \includegraphics[scale=0.5,angle=0]{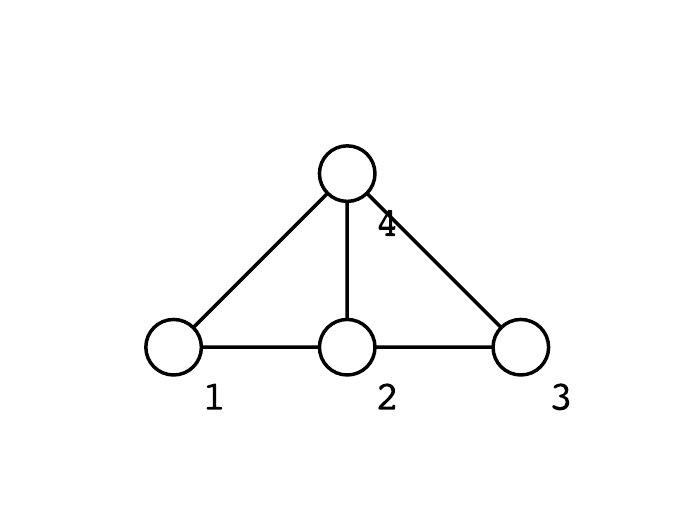}  \quad \includegraphics[scale=0.5,angle=0]{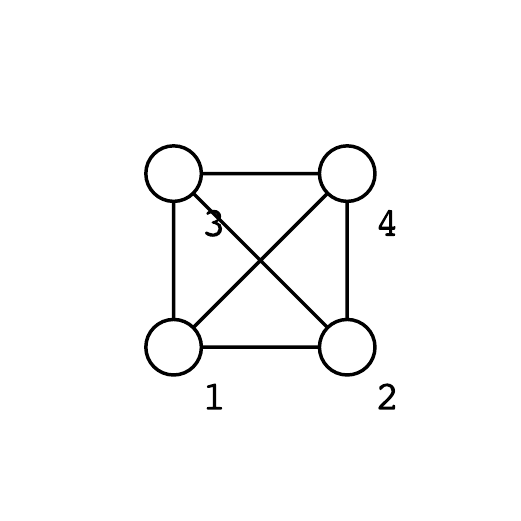}
\end{align}
The first four are the Dynkin diagrams of $SL(5,\mathbb{R})$, $SO(8)$, $SL(3,\mathbb{R})^{++}$ and $SL(4,\mathbb{R})^+$, the final two are Dynkin diagrams of indefinite type. We concentrate on examples of the first two types of Dynkin diagram although we will construct the real form $SO(4,4)$ of $SO(8)$ in all our examples which is identified using the temporal involution in each example that will be discussed.

\subsubsection*{An $\frac{SL(5,\mathbb{R})}{SO(2,3)}$ bound state of $(F1,D2,D4,NS5,D6,KK5)$.}
There are many bound states having an $SL(5,\mathbb{R})$ symmetry, however most of these bound states include exotic (or non-geometric) branes with an $SL(5,\mathbb{R})$ symmetry - these objects give rise to brane solutions upon dimensional reduction. Here we consider an example which includes a KK5 brane, or gravitational monopole, from which the D5-brane in nine-dimensions is derived. 

In table \ref{rank_4_sl_5_roots} are listed four branes whose associated roots in the root system of $E_{11}$ are the simple positive roots of $SL(5,\mathbb{R})$, while in table \ref{rank_4_sl_5_branes} we have listed the orientation of the D-branes involved in the bound state.
\begin{table}[ht] 
\centering 
\scalebox {0.95} {
\begin{tabular}{ c | c | c | c | c | c | c | c | c | c | c  } 
Simple Roots &1& 2& 3& 4& 5& 6& 7& 8& 9(t)& 10  \\
\hline
S0&&&&&&&&&&$\bullet$   \\
\hline 
F1&&&&&&&&$\bullet$ &$\bullet$ &  \\
\hline 
S2&&&&&$\bullet$ &$\bullet$ &$\bullet$ &&&  \\
\hline 
S0&&&&$\bullet$ &&&&&&$\bullet$  
\end{tabular} }
\caption{An example of the simple roots of $\mathfrak{sl}(5,\mathbb{R})$ described as oriented branes.}\label{rank_4_sl_5_roots}
\end{table}
\begin{table}[ht] 
\centering 
\scalebox {0.95} {
\begin{tabular}{ c | c | c | c | c | c | c | c | c | c | c  } 
Simple Roots &1& 2& 3& 4& 5& 6& 7& 8& 9(t)& 10  \\
\hline 
F1&&&&&&&&$\bullet$ &$\bullet$ &  \\
\hline 
D2&&&&&&&&$\bullet$&$\bullet$&$\bullet$  \\
\hline 
D4&&&&&$\bullet$&$\bullet$&$\bullet$&$\bullet$&$\bullet$&  \\
\hline 
NS5&&&&&$\bullet$&$\bullet$&$\bullet$&$\bullet$&$\bullet$&$\bullet$  \\
\hline 
D6&&&&$\bullet$&$\bullet$&$\bullet$&$\bullet$&$\bullet$&$\bullet$&$\bullet$  \\
\hline 
KK5&&&&$\bullet$&$\bullet$&$\bullet$&$\bullet$&$\bullet$&$\bullet$&$\odot$  \\
\end{tabular} }
\caption{An example of the branes present in an $SL(5,\mathbb{R})$ bound state. Where $\odot$ indicates the NUT direction.}\label{rank_4_sl_5_branes}
\end{table}

The Borel sub-algebra of $\mathfrak{sl}(5,\mathbb{R})$ is embedded in $\mathfrak{e}_{11}$ as follows
\begin{align}
\nonumber H_1&=-\frac{1}{8}({K^1}_1+\ldots + {K^9}_9)+\frac{7}{8}{K^{10}}_{10}-\frac{3}{2}{R},\\
\nonumber H_2&=-\frac{1}{4}({K^1}_1+\ldots +{K^7}_7+{K^{10}}_{10})+\frac{3}{4}({K^8}_8+{K^9}_9)+R,\\
\nonumber H_3&=-\frac{3}{8}({K^1}_1+\ldots+ {K^4}_4+{K^8}_8+\ldots +{K^{10}}_{10})\\&\qquad \qquad +\frac{5}{8}({K^5}_5+{K^6}_6 +{K^7}_7)-\frac{1}{2}R,\\
\nonumber H_4&=-\frac{1}{4}({K^1}_1+\ldots +{K^3}_3+{K^5}_5+\ldots +{K^{9}}_{9})+\frac{3}{4}({K^4}_4+{K^{10}}_{10})+R,\\
\nonumber E_1&=R^{10}, \quad E_2=R^{89}, \quad E_{3}=R^{567}, \quad E_{4}=R^{410},\\
\nonumber E_{12}&=R^{8910}, \quad E_{23}=R^{56789}, \quad E_{34}=-R^{456710}, \\
\nonumber E_{123}&=-R^{5678910}, \quad E_{234}=-R^{45678910} \quad \mbox{and}\\
\nonumber E_{1234}&=-R^{45678910,10}.
\end{align}
The non-zero commutators of the positive generators are
\begin{align}
\nonumber [E_1,E_2]&=E_{12}, \quad [E_2,E_3]=E_{23}, \quad [E_3,E_4]=E_{34}, \\
 [E_{12},E_{3}]&=[E_1,E_{23}]=E_{123},\quad [E_{2},E_{34}]=[E_{23},E_4]=E_{234} \quad \mbox{and}\\
\nonumber [E_1,E_{234}]&=[E_{12},E_{34}]=[E_{123},E_4]=E_{1234}.
\end{align}
In this example we will treat $x^9$ as the temporal coordinate, so that the generators of the simple positive roots $(E_1,E_2,E_3,E_4)$ are associated to the gauge fields of (S0, F1, S2, S0) string theory solutions, as indicated in table \ref{rank_4_sl_5_roots}. This choice corresponds to the temporal involution
\begin{equation}
\Omega(E_1)=-F_1,\quad \Omega(E_2)=F_2, \quad \Omega(E_3)=-F_3,\quad \Omega(E_4)=-F_4
\end{equation}
and the sub-algebra invariant under $\Omega$ is $\mathfrak{so}(2,3)$. The coset model of $SL(5,\mathbb{R})$ has not been studied in the literature before and so we will cover it in more detail than the previous examples. 
The representative coset element is:
\begin{align}
\nonumber g= \exp(&\phi_1(\xi)H_1+\phi_2(\xi)H_2+\phi_3(\xi)H_3+\phi_4(\xi)H_4)\exp(C_1(\xi)E_1+C_2(\xi)E_2\\
\nonumber &+C_3(\xi)E_3+C_4(\xi)E_4+C_{12}(\xi)E_{12}+C_{23}(\xi)E_{23}+C_{34}(\xi)E_{34}\\
&+C_{123}(\xi)E_{123}+C_{234}(\xi)E_{234}+C_{1234}(\xi)E_{1234}).\label{sl5groupelement}
\end{align}
The component of the Maurer-Cartan form can then be written as
\begin{align}
\nonumber \partial g g^{-1}=&\partial\phi_1H_1+\partial\phi_2H_2+\partial\phi_3H_3+\partial\phi_4H_4+P_1S_1+P_2S_2+P_3S_3\\
&+P_4S_4+P_{12}S_{12}+P_{23}S_{23}+P_{34}S_{34}+P_{123}S_{123}+P_{234}S_{234}\\
\nonumber &+P_{1234}S_{1234}
\end{align}
where $H_i$ and $S_i$ are the remaining generators in the Borel sub-algebra of $\mathfrak{sl}(5,\mathbb{R})$ after the $\mathfrak{so}(2,3)$ sub-algebra has been removed. The $P_i$ are
\begin{align}
P_1&=\partial C_1 \exp{(2\phi_1-\phi_2)} \label{Cequation1}\\
P_2&=\partial C_2 \exp{(-\phi_1+2\phi_2-\phi_3)}\\
P_3&=\partial C_3 \exp{(-\phi_2+2\phi_3-\phi_4)}\\
P_4&=\partial C_4 \exp{(-\phi_3+2\phi_4)}\\
P_{12}&=(\partial C_{12}-\frac{1}{2}(\partial C_1 C_2 -\partial C_2 C_1))\exp{(\phi_1+\phi_2-\phi_3)}\\
P_{23}&=(\partial C_{23}-\frac{1}{2}(\partial C_2 C_3 -\partial C_3 C_2))\exp{(-\phi_1+\phi_2+\phi_3-\phi_4)}\\
P_{34}&=(\partial C_{34}-\frac{1}{2}(\partial C_3 C_4 -\partial C_4 C_3))\exp{(-\phi_2+\phi_3+\phi_4)}\\
P_{123}&=(\partial C_{123}-\frac{1}{2}(\partial C_1 C_{23} -\partial C_3 C_{12}+\partial C_{12} C_3-\partial C_{23}C_1)\\
\nonumber &\quad +\frac{1}{3!}(\partial C_1 C_2 C_3-2\partial C_2 C_1 C_3+\partial C_3 C_2 C_1)\exp{(\phi_1+\phi_3-\phi_4)}\\
P_{234}&=(\partial C_{234}-\frac{1}{2}(\partial C_2 C_{34} -\partial C_4 C_{23}+\partial C_{23} C_4-\partial C_{34}C_2)\\
\nonumber &\quad +\frac{1}{3!}(\partial C_2 C_3 C_4-2\partial C_3 C_2 C_4+\partial C_4 C_3 C_2)\exp{(-\phi_1+\phi_2+\phi_4)}\\
P_{1234}&=(\partial C_{1234}-\frac{1}{2}(\partial C_1 C_{234} -\partial C_4 C_{123}+\partial C_{12} C_{34}-\partial C_{34}C_{12} \label{Cequation14}\\
\nonumber & \quad+\partial C_{123} C_{4}-\partial C_{234}C_{1})+\frac{1}{3!}(\partial C_1 C_2 C_{34}-2\partial C_2 C_1 C_{34}\\
\nonumber &\quad +\partial C_{34} C_2 C_1+\partial C_1 C_{23} C_4-2\partial C_{23} C_1 C_4+\partial C_4 C_{23} C_1\\
\nonumber &\quad + \partial C_{12} C_3 C_4-2\partial C_3 C_{12} C_4+\partial C_4 C_{12} C_3)-\frac{1}{4!}(\partial C_1C_2C_3C_4\\
\nonumber &\quad - 3\partial C_2 C_1C_3C_4+ 3\partial C_3 C_1C_2C_4-\partial C_4 C_1C_2C_3))\exp{(\phi_1+\phi_4)}
\end{align}
The equation of motion (\ref{eom}) gives
\begin{align}
\partial^2\phi_1+\frac{1}{2}P_1^2+\frac{1}{2}P_{12}^2+\frac{1}{2}P_{123}^2-\frac{1}{2}P_{1234}^2&=0\label{1}\\
\partial^2\phi_2-\frac{1}{2}P_2^2+\frac{1}{2}P_{12}^2-\frac{1}{2}P_{23}^2+\frac{1}{2}P_{123}^2+\frac{1}{2}P_{234}^2-\frac{1}{2}P_{1234}^2&=0\label{2}\\
\partial^2\phi_3-\frac{1}{2}P_3^2-\frac{1}{2}P_{23}^2+\frac{1}{2}P_{34}^2+\frac{1}{2}P_{123}^2+\frac{1}{2}P_{234}^2-\frac{1}{2}P_{1234}^2&=0\label{3}\\
\partial^2\phi_4+\frac{1}{2}P_4^2+\frac{1}{2}P_{34}^2+\frac{1}{2}P_{234}^2-\frac{1}{2}P_{1234}^2&=0\label{4}\\
\partial P_1+(2\partial \phi_1-\partial \phi_2)P_1-P_2P_{12}-P_{23}P_{123}+P_{234}P_{1234}&=0\label{5}\\
\partial P_2+(-\partial \phi_1+2\partial \phi_2-\partial\phi_3)P_2-P_1P_{12}-P_{3}P_{23}+P_{34}P_{234}&=0\label{6}\\
\partial P_3+(-\partial \phi_2+2\partial \phi_3-\partial\phi_4)P_3+P_2P_{23}+P_{4}P_{34}-P_{12}P_{123}&=0\label{7}\\
\partial P_4+(-\partial \phi_3+2\partial \phi_4)P_4+P_3P_{34}+P_{23}P_{234}-P_{123}P_{1234}&=0\label{8}\\
\partial P_{12}+(\partial \phi_1+\partial \phi_2-\partial\phi_3)P_{12}-P_3P_{123}+P_{34}P_{1234}&=0\label{9}\\
\partial P_{23}+(-\partial \phi_1+\partial \phi_2+\partial\phi_3-\partial\phi_4)P_{23}-P_1P_{123}+P_{4}P_{234}&=0\label{10}\\
\partial P_{34}+(-\partial \phi_2+\partial \phi_3+\partial\phi_4)P_{34}+P_2P_{234}-P_{12}P_{1234}&=0\label{11}\\
\partial P_{123}+(\partial \phi_1+\partial \phi_3-\partial\phi_4)P_{123}+P_4P_{1234}&=0\label{12}\\
\partial P_{234}+(-\partial \phi_1+\partial \phi_2+\partial\phi_4)P_{234}-P_1P_{1234}&=0\label{13}\\
\partial P_{1234}+(\partial \phi_1+\partial \phi_4)P_{1234}&=0 \label{14}
\end{align}
These equations, as well as quadratic Hamiltonian constraint $(P_\xi|P_\xi)=0$ are solved by 
\begin{align}
\nonumber &\qquad \phi_1=\frac{1}{2}\ln N_1,\quad \phi_2=\frac{1}{2}\ln N_2,\quad \phi_3=\frac{1}{2}\ln N_3, \quad \phi_4=\frac{1}{2}\ln N_4,\\
\nonumber P_1&=\sqrt{\frac{\alpha_{12}}{\partial N_1}}\frac{\partial N_1}{N_1\sqrt{N_2}}, P_2=\frac{\sqrt{\alpha_{12}\alpha_{23}}}{N_2\sqrt{N_1N_3}},P_3=\frac{\sqrt{\alpha_{23}\alpha_{34}}}{N_3\sqrt{N_2N_4}},P_4=\sqrt{\frac{\alpha_{34}}{\partial N_4}}\frac{\partial N_4}{N_4\sqrt{N_3}},\\
\nonumber &\qquad P_{12}=\frac{\sqrt{\partial N_1\alpha_{23}}}{\sqrt{N_1N_2N_3}}, P_{23}=-\frac{\sqrt{\alpha_{12}\alpha_{34}}}{\sqrt{N_1N_2N_3N_4}}, P_{34}=-\frac{\sqrt{\partial N_4\alpha_{23}}}{\sqrt{N_2N_3N_4}},\\
&\qquad\qquad \qquad P_{123}=-\frac{\sqrt{\partial N_1\alpha_{34}}}{\sqrt{N_1N_3N_4}}, \qquad P_{234}=-\frac{\sqrt{\partial N_4\alpha_{12}}}{\sqrt{N_1N_2N_4}},\\
\nonumber &\qquad \qquad \qquad \qquad \qquad \quad P_{1234}=\frac{\sqrt{\partial N_1\partial N_4}}{\sqrt{N_1N_4}}
\end{align}
where the $N_i$ are harmonic functions in one dimension given by
\begin{align}
\nonumber N_1&=1+q\xi\cos^2{\alpha},\\
N_2&=1+q\xi,\\
\nonumber N_3&=1+q\xi\cos^2{\beta} \quad \mbox{and}\\
\nonumber N_4&=1+q\xi\cos^2{\beta}\cos^2{\gamma}
\end{align}
and 
\begin{align}
\nonumber \alpha_{12}&\equiv N_1\partial N_2 - N_2\partial N_1=q\sin^2{\alpha},\\
\alpha_{23}&\equiv N_3\partial N_2 - N_2\partial N_3=q\sin^2{\beta}\quad \mbox{and}\\
\nonumber \alpha_{34}&\equiv N_4\partial N_3 - N_3\partial N_4=q\cos^2{\beta}\sin^2{\gamma}.
\end{align}
The $C$ fields we commenced with in (\ref{sl5groupelement}) and are found by solving 
equations (\ref{Cequation1}-\ref{Cequation14}) giving
\begin{align}
\nonumber C_1&=-\sqrt{\frac{\alpha_{12}}{\partial N_1}}\frac{1}{N_1}, \quad C_2=-\frac{\sqrt{\alpha_{12}\alpha_{23}}}{\partial N_2}\frac{1}{N_2}, \quad C_3=-\frac{\sqrt{\alpha_{23}\alpha_{34}}}{\partial N_3}\frac{1}{N_3},\\
\nonumber C_4&=-\sqrt{\frac{\alpha_{34}}{\partial N_4}}\frac{1}{N_4},\quad C_{12}=-\frac{\sqrt{\partial N_1 \alpha_{23}}}{2\partial N_1 \partial N_2}\left(\frac{\partial N_1}{N_1}+\frac{\partial N_2}{N_2}\right),\\
\nonumber  C_{23}&=\frac{\sqrt{\alpha_{12}\alpha_{34}}}{2\partial N_2 \partial N_3}\left(\frac{\partial N_2}{N_2}+\frac{\partial N_3}{N_3}\right),\quad C_{34}=\frac{\sqrt{\alpha_{23}\partial N_4}}{2\partial N_3 \partial N_4}\left(\frac{\partial N_3}{N_3}+\frac{\partial N_4}{N_4}\right),\\
C_{123}&=\sqrt{\frac{\alpha_{34}}{\partial N_1}}\frac{1}{\partial N_3}\left(\frac{\partial N_1}{3N_1}+\frac{\partial N_2}{6N_2}+\frac{\partial N_3}{3N_3}+\frac{N_2\partial N_1\partial N_3}{6N_1N_3\partial N_2}\right),\\
\nonumber  C_{234}&=-\sqrt{\frac{\alpha_{12}}{\partial N_4}}\frac{1}{\partial N_2}\left(\frac{\partial N_2}{3N_1}+\frac{\partial N_3}{6N_3}+\frac{\partial N_4}{3N_4}+\frac{N_3\partial N_2\partial N_4}{6N_2N_4\partial N_3}\right) \quad \mbox{and}\\
\nonumber C_{1234}&=-\frac{1}{\sqrt{\partial N_1\partial N_4}}\bigg(\frac{\partial N_1}{4N_1}+\frac{\partial N_2}{12N_2}+\frac{\partial N_3}{12N_3}+\frac{\partial N_4}{4N_4}\\
\nonumber & \qquad +\frac{N_3\partial N_4}{12N_4\partial N_3}\bigg(\frac{\partial N_1}{N_1}+\frac{\partial N_2}{N_2}\bigg) +\frac{N_2\partial N_1}{12N_1\partial N_2}\bigg(\frac{\partial N_3}{N_3}+\frac{\partial N_4}{N_4}\bigg) \bigg).
\end{align}
The diagonal components of the metric\footnote{There are also non-zero off-diagonal components of the metric in this example due to the $KK5$ monopole, but we shall first need the diagonal metric to derive the off-diagonal contribution.} are read from the coefficients of the Cartan sub-algebra when the solution is substituted in (\ref{sl5groupelement}) and are:
\begin{align}
ds^2&=N_1^\frac{1}{8}N_2^\frac{1}{4}N_3^\frac{3}{8}N_4^\frac{1}{4}(\sum_{i=1}^{3}(dx^i)^2+N_4^{-1}(dx^4)^2+N_3^{-1}\sum_{i=5}^{7}(dx^i)^2 \label{sl5metric}\\
\nonumber &\quad +N_2^{-1}((dx^8)^2-(dt^9)^2)+N_1^{-1}N_4^{-1}(dx^{10})^2)
\end{align}
and the dilaton is
\begin{equation}
e^A=N_1^\frac{3}{4}N_2^{-\frac{1}{2}}N_3^\frac{1}{4}N_4^{-\frac{1}{2}}.
\end{equation}
By identifying the world-line parameter $\xi$ to the space-time coordinate $x^1$ (any of the directions transverse to all component branes would suffice equally well) we may explicitly write the active field strength components for the solution. To do this we make use of the commutators:
\begin{equation}
[R,E_i]\equiv a_iE_i.
\end{equation}
For the generators in this solution we find
\begin{align}
\nonumber a_1&=-\frac{3}{4}, \quad a_2=\frac{1}{2},\quad a_3=-\frac{1}{4},\quad a_4=\frac{1}{2},\\
a_{12}&=-\frac{1}{4},\quad a_{23}=\frac{1}{4}, \quad a_{34}=\frac{1}{4},\\
\nonumber a_{123}&=-\frac{1}{2},\quad a_{234}=\frac{3}{4} \quad \mbox{and}\\
\nonumber a_{1234}&=0.
\end{align}
The field strength components may now be derived using $F_i=(e^A)^{a_i}P_i$ and embedding the tensor indices in space-time using the vielbein encoded in the diagonal metric in (\ref{sl5metric}). The results are
\begin{align}
G&=-\tan{\alpha}\partial N_1^{-1}  dx^1\wedge dx^{10} -\sin{\alpha}\cos{\beta}\cos{\gamma} \frac{e^{\frac{3A}{2}}\partial N_2}{N_1} dx^2\wedge dx^3,\\
H&=-\sin{\alpha}\sin{\beta}\partial N_2^{-1} dx^1\wedge dx^8\wedge dt^9-\tan{\gamma}\partial N_4^{-1}dx^1\wedge dx^4\wedge dx^{10}\\
\nonumber &\quad -\cos{\alpha}\cos{\beta}\cos{\gamma} \frac{e^{-A}\partial N_2}{N_4} dx^2\wedge dx^3 \wedge dx^{10}\quad \mbox{and}\\
F&=-\tan{\beta}\sin{\gamma}\partial N_3^{-1}dx^1\wedge dx^5 \wedge dx^6 \wedge dx^7\\
\nonumber &\quad +\cos{\alpha}\sin{\beta}\frac{\partial N_2}{N_1N_2} dx^1\wedge dx^8\wedge dt^9\wedge dx^{10}\\
\nonumber &\quad +\sin{\alpha}\cos{\beta}\sin{\gamma}\frac{e^{\frac{A}{2}}\partial N_2}{N_1N_4}  dx^2\wedge dx^3 \wedge dx^4 \wedge dx^{10}\\
\nonumber &\quad +\cos{\beta}\sin{\beta}\cos{\gamma} \frac{e^{\frac{A}{2}}\partial N_2}{N_2} dx^2\wedge dx^3 \wedge dx^8 \wedge dt^{9}.
\end{align}
Finally we consider the field strength associated to the KK5 monopole. Proceeding in the same manner as the other field strengths we find
\begin{equation}
F^{KK}=-\cos{\alpha}\cos{\beta}\cos{\gamma}\frac{\partial N_2}{N_1N_4}dx^1\wedge dx^4\wedge dx^5 \wedge \ldots \wedge dx^{10}\otimes dx^{10}
\end{equation}
Upon dualisation and the raising of the $x^{10}$ index the non-zero component is
\begin{equation}
{(*F^{KK})_{23}}^{10}=\cos{\alpha}\cos{\beta}\cos{\gamma}\partial N_2.
\end{equation}
It is useful to localise the solution by upgrading the harmonic functions in $\xi=x^1$ to harmonic functions in $r\equiv \sqrt{(x^1)^2+(x^2)^2+(x^3)^2}$. By changing to spherical coordinates $r,\theta,\phi$ in the transverse three dimensional space and solving the monopole equation $dA=\nabla N_2 \cos{\alpha}\cos{\beta}\cos{\gamma}$ we find the full metric for the solution is given by
\begin{align}
\nonumber ds^2&=N_1^\frac{1}{8}N_2^\frac{1}{4}N_3^\frac{3}{8}N_4^\frac{1}{4}\bigg(dr^2+r^2d\theta^2+r^2\sin^2\theta d\phi^2+N_4^{-1}(dx^4)^2\\
&\qquad +N_3^{-1}((dx^5)^2+(dx^6)^2+(dx^7)^2) +N_2^{-1}((dx^8)^2-(dt^9)^2)\\
\nonumber &\qquad+N_1^{-1}N_4^{-1}(dx^{10}+\partial N_2 \cos{\alpha}\cos{\beta}\cos{\gamma}\cos\theta d\phi)^2\bigg).
\end{align}
The conical singularity in the limit $r\rightarrow 0$ and $\alpha=\beta=\gamma=0$ is avoided if $x^{10}$ is cyclic with period $2\pi$, however the asymptotic topology of this solution as the parameters $\alpha,\beta,\gamma$ vary deserves further study as the solution encodes a topology change of spacetime.
\subsubsection*{Comments on $SO(4,4)$ bound states.}
Up to the choice of the real form of the local sub-algebra, $\cal K$, there are four distinct bound states possessing a global $SO(4,4)$ symmetry which are all formed of D-branes and one KK-brane. Apart from the exception the KK-brane is the KK5 brane, which is the ten-dimensional dual graviton, these bound states require dimensional reduction before they may be interpreted as pure bound states. The algebra of $SO(p,q)$ is sufficiently different to the examples considered previously that it merits its own detailed discussion and will be presented elsewhere. 

We will first present the four cases where $x^{10}$ is the temporal direction and latterly we will discuss alternative choices of the temporal coordinate which determines $\cal K$. Of the four cases the first is a bound state of a D0 brane, three D2 branes, three D4 branes and a D6 brane, oriented as shown in table \ref{rank_4_so_8_1}, which we will indicate by (D0,D2$^3$,D4$^3$,D6)\footnote{Where we indicate the multiplicity of the branes involved by the superscript number.}. This state was discovered by considering the deformations of the D6 brane in \cite{Larsson:2001p2275}, but that it may be described as a one-dimensional $\sigma$-model on a coset of $SO(4,4)$ was not known. Additionally we find bound states of (F1,D2,D4$^2$,D6,KK5), (F1,D4,D6$^2$,D8,KK4$_3$) and (D2,D4$^3$,D6$^3$,D8) which also have a global $SO(4,4)$ symmetry. The orientations of the branes in these states are shown in tables \ref{rank_4_so_8_2}, \ref{rank_4_so_8_3} and \ref{rank_4_so_8_4}.
\begin{table}[ht] 
\centering 
\scalebox {0.95} {
\begin{tabular}{ c | c | c | c | c | c | c | c | c | c | c  } 
Branes &1& 2& 3& 4& 5& 6& 7& 8& 9& 10(t)  \\
\hline
D0&&&&&&&&&&$\bullet$   \\
\hline 
D2&&&&&&&&$\bullet$ &$\bullet$ &$\bullet$   \\
\hline 
D2&&&&&&$\bullet$ &$\bullet$ &&&$\bullet$   \\
\hline 
D2&&&&$\bullet$ &$\bullet$ &&&&&$\bullet$   \\
\hline 
D4&&&&&&$\bullet$ &$\bullet$ &$\bullet$ &$\bullet$ &$\bullet$   \\
\hline 
D4&&&&$\bullet$ &$\bullet$ &&&$\bullet$ &$\bullet$ &$\bullet$   \\
\hline 
D4&&&&$\bullet$ &$\bullet$ &$\bullet$ &$\bullet$ &&&$\bullet$   \\
\hline 
D6&&&&$\bullet$ &$\bullet$ &$\bullet$ &$\bullet$ &$\bullet$ &$\bullet$ &$\bullet$   \\
\end{tabular} }
\caption{The $SO(4,4)$ bound states of (D0,D2$^3$,D4$^3$,D6)}\label{rank_4_so_8_1}
\end{table}
\begin{table}[ht] 
\centering 
\scalebox {0.95} {
\begin{tabular}{ c | c | c | c | c | c | c | c | c | c | c  } 
Branes &1& 2& 3& 4& 5& 6& 7& 8& 9& 10(t)  \\
\hline
F1&&&&&&&&&$\bullet$ &$\bullet$   \\
\hline 
D2&&&&&&&&$\bullet$ &$\bullet$ &$\bullet$   \\
\hline 
D4&&&&&&$\bullet$ &$\bullet$ &$\bullet$ &$\bullet$ &$\bullet$   \\
\hline 
D4&&&&$\bullet$ &$\bullet$ &&&$\bullet$ &$\bullet$ &$\bullet$   \\
\hline 
D6&&&&$\bullet$ &$\bullet$ &$\bullet$ &$\bullet$ &$\bullet$ &$\bullet$ &$\bullet$   \\
\hline 
KK5&&&&$\bullet$ &$\bullet$ &$\bullet$ &$\bullet$ &$\odot$ &$\bullet$ &$\bullet$   \\
\end{tabular} }
\caption{The $SO(4,4)$ bound states of (F1,D2,D4$^2$,D6,KK5)}\label{rank_4_so_8_2}
\end{table}
\begin{table}[ht] 
\centering 
\scalebox {0.95} {
\begin{tabular}{ c | c | c | c | c | c | c | c | c | c | c  } 
Branes &1& 2& 3& 4& 5& 6& 7& 8& 9& 10(t)  \\
\hline
F1&&&&&&&&&$\bullet$ &$\bullet$   \\
\hline 
D4&&&&&&$\bullet$ &$\bullet$ &$\bullet$ &$\bullet$ &$\bullet$   \\
\hline 
D6&&&&$\bullet$ &$\bullet$ &$\bullet$ &$\bullet$ &$\bullet$ &$\bullet$ &$\bullet$   \\
\hline 
D6&&$\bullet$ &$\bullet$ &&&$\bullet$ &$\bullet$ &$\bullet$ &$\bullet$ &$\bullet$   \\
\hline 
D8&&$\bullet$ &$\bullet$ &$\bullet$ &$\bullet$ &$\bullet$ &$\bullet$ &$\bullet$ &$\bullet$ &$\bullet$   \\
\hline 
KK4$_3$&&$\bullet$ &$\bullet$ &$\bullet$ &$\bullet$ &$\odot$ &$\odot$ &$\odot$ &$\bullet$ &$\bullet$   \\
\end{tabular} }
\caption{The $SO(4,4)$ bound states of (F1,D4,D6$^2$,D8)}\label{rank_4_so_8_3}
\end{table}
\begin{table}[ht] 
\centering 
\scalebox {0.95} {
\begin{tabular}{ c | c | c | c | c | c | c | c | c | c | c  } 
Branes &1& 2& 3& 4& 5& 6& 7& 8& 9& 10(t)  \\
\hline
D2&&&&&&&&$\bullet$ &$\bullet$ &$\bullet$   \\
\hline 
D4&&&&&&$\bullet$ &$\bullet$ &$\bullet$ &$\bullet$ &$\bullet$   \\
\hline 
D4&&&&$\bullet$ &$\bullet$ &&&$\bullet$ &$\bullet$ &$\bullet$   \\
\hline 
D4&&$\bullet$ &$\bullet$ &&&&&$\bullet$ &$\bullet$ &$\bullet$   \\
\hline 
D6&&&&$\bullet$ &$\bullet$ &$\bullet$ &$\bullet$ &$\bullet$ &$\bullet$ &$\bullet$   \\
\hline 
D6&&$\bullet$ &$\bullet$ &&&$\bullet$ &$\bullet$ &$\bullet$ &$\bullet$ &$\bullet$   \\
\hline 
D6&&$\bullet$ &$\bullet$ &$\bullet$ &$\bullet$ &&&$\bullet$ &$\bullet$ &$\bullet$   \\
\hline 
D8&&$\bullet$ &$\bullet$ &$\bullet$ &$\bullet$ &$\bullet$ &$\bullet$ &$\bullet$ &$\bullet$ &$\bullet$   \\
\end{tabular} }
\caption{The $SO(4,4)$ bound states of (D2,D4$^3$,D6$^3$,D8)}\label{rank_4_so_8_4}
\end{table}

It is anticipated that the full solutions will be understood by a null geodesic on a coset of $SO(4,4)$. As in the previous examples the choice of temporal involution is given by the embedding of the algebra in $E_{11}$ and depends upon which space-time coordinates are temporal. Consider the example of the bound state listed in table \ref{rank_4_so_8_1}. It consists of the following generators embedded in $E_{11}$
\begin{align}
\nonumber E_{1}&=R^{89}, \quad E_{2}=R^{10}, \quad E_{3}=R^{67}, \quad E_{4}=R^{45}, \\
\nonumber E_{12}&=R^{8910}, \quad E_{23}=R^{6710}, \quad E_{24}=R^{4510},\\
E_{123}&=R^{678910}, \quad E_{124}=R^{458910}, \quad E_{234}=R^{456710},\\
\nonumber E_{1234}&=R^{45678910} \quad \mbox{and}\\
\nonumber \quad E_{21234}&=R^{45678910,10}.
\end{align}
The subscript on the generators indicates, up to a sign, the non-trivial commutation relations (e.g. $E_{123}=[E_1,E_{23}]=[E_1,E_2,E_3]$). In the example $x^{10}$ was chosen to be a temporal coordinate and hence the temporal involution on these generators follows from the temporal involution of $E_{11}$ which selects $x^{10}$ to be timelike, i.e.
\begin{equation}
\Omega(E_1)=-F_1, \; \Omega(E_2)=F_2, \; \Omega(E_3)=-F_3 \; \mbox{ and } \;  \Omega(E_4)=-F_4.
\end{equation}
The local sub-algebra is invariant under the temporal involution and hence it contains 4 compact generators and 8 non-compact generators. The only sub-group of $SO(4,4)$ whose algebra contains these numbers of compact and non-compact generators is $SO(2,2)\times SO(2,2)$. This is schematically indicated in figure \ref{realforms}(a.). The example indicated in table \ref{rank_4_so_8_4} is of the same type and its solution is also expected to be described by a null geodesic on a coset of $\frac{SO(4,4)}{SO(2,2)\times SO(2,2)}$.
\begin{figure}
\centering
\subfloat[${\cal K}=SO(2,2)\times SO(2,2)$]{\hspace{35pt} \includegraphics[scale=0.65,angle=0]{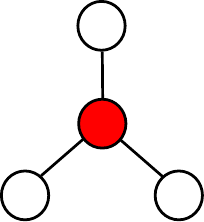}\hspace{35pt}}
\hspace{40pt}
\subfloat[${\cal K}=SO(1,3)\times SO(1,3)$]{\hspace{35pt} \includegraphics[scale=0.65,angle=0]{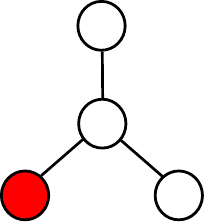}\hspace{35pt}} 
\caption{The real form of the local sub-group ${\cal K}$ of $SO(4,4)$ when an electric brane is associated to the shaded node.} \label{realforms}
\end{figure}

The remaining two bound states indicated in tables \ref{rank_4_so_8_2} and \ref{rank_4_so_8_3} have the local sub-group $SO(1,3)\times SO(1,3)$ chosen as indicated in figure \ref{realforms}(b.). To see this consider the example of table \ref{rank_4_so_8_3} for which the embedding in the generators of $\mathfrak{e}_{11}$ is 
\begin{align}
\nonumber E_{1}&=R^{910}, \quad E_{2}=R^{8}, \quad E_{3}=R^{67}, \quad E_{4}=R^{45}, \\
\nonumber E_{12}&=R^{8910}, \quad E_{23}=R^{678}, \quad E_{24}=R^{458},\\
E_{123}&=R^{678910}, \quad E_{124}=R^{458910}, \quad E_{234}=R^{45678},\\
\nonumber E_{1234}&=R^{45678910} \quad \mbox{and}\\
\nonumber \quad E_{21234}&=R^{45678910,8}.
\end{align}
As $x^{10}$ is temporal the temporal involution acts on the generators as
\begin{equation}
\Omega(E_1)=F_1, \; \Omega(E_2)=-F_2, \; \Omega(E_3)=-F_3 \; \mbox{ and } \;  \Omega(E_4)=-F_4.
\end{equation}
Consequently the involution-invariant sub-algebra consists of 6 non-compact generators and 6 compact generators which is uniquely matched within the real form of $\mathfrak{so}(8)$ by the algebra of the group $SO(1,3)\times SO(1,3)$. The construction of the solution requires an ansatz different to the $SL(n,\mathbb{R})$ examples considered previously and we leave the detailed discussion of the null geodesic motion on $\frac{SO(4,4)}{SO(2,2)\times SO(2,2)}$ and $\frac{SO(4,4)}{SO(1,3)\times SO(1,3)}$ to be presented elsewhere.
\end{subsubsection}

\begin{subsection}{Low rank IIB string theory bound states.}
As for the case of type-IIA string theory we aim to identify all the recognisable Cartan matrices $A_{ij}=<\beta_i,\beta_j>$ where $\beta_i$  are roots of $E_{11}$ associated to IIB D-branes, NS-branes and S-branes. Specifically we will consider bound states formed of the following branes and their Euclidean (S-brane) counterparts: the D1, F1, D3, D5, NS5, D7$_a$, D7$_b$, D9$_a$ and D9$_b$ branes. The derivation of the roots associated to these branes is given in section 2.2 and their Young tableaux, from which the usual root expansion may be read, is shown in table \ref{table:IIB}. We carry out the same process as described for the type-IIA theory and again will concern ourselves only with solutions described as cosets of finite groups whose Dynkin diagrams are simply-laced. In particular at low orders we will highlight the bound states devoid of exotic content.

\begin{subsubsection}{Type-IIB: Cosets of rank two groups.}
Within the roots of $E_{11}$ decomposed into the IIB representations there are seventy-four bound states whose solutions are described by a null geodesic on the coset $\frac{SL(3,\mathbb{R})}{SO(1,2)}$. Of these only fourteen bound states involve only roots associated to standard type-IIB solutions and we list them for reference in table \ref{IIB_rank_2_bound_states}. 
\begin{table}[ht]
\tabcolsep 3pt
\small
\centering
\begin{tabular}{ c || c | c | c | c | c | c | c }
& S1$_a$ & S1$_b$ & S3 & S5$_b$ & S5$_a$ & S7$_a$ & S7$_b$ \\
\hline
\hline
F1& - & (F1,D3) & (F1,D5) & - & (F1,D7$_a$) & (F1,D9$_a$) & -  \\
\hline
D1 & (D1,D3) & - & (D1,NS5) & (D1,D7$_b$) & - & - & (D1,D9$_b$)  \\
\hline
D3 & (D3,D5) & (D3,NS5) & - & - & - & - & -   \\
\hline
NS5 & - & (NS5,D7$_b$) & - & - & - & -  & -  \\
\hline
D5 & (D5,D7$_a$) & - & - & - & - & - & -  \\
\hline
D7$_a$ & (D7$_a$,D9$_a$) & - & - & - & - & - & -   \\
\hline
D7$_b$ & - & (D7$_b$,D9$_b$)  & - & - & - & - & -   \\
\end{tabular}
\caption{The $SL(3,\mathbb{R})$ bound states of the canonical IIB branes. In particular, S1$_a$ indicates a spacelike fundamental string, S1$_b$ indicates a spacelike D1 brane, S5$_a$ indicates a spacelike D5 and S5$_b$ indicates a spacelike NS5.} \label{IIB_rank_2_bound_states}
\end{table}
Each of the states shown in table \ref{IIB_rank_2_bound_states} has a solution which may be explicitly found using the techniques of \cite{Houart:2010p2928} and a solution algebraically identical is given there and reproduced here in equations (\ref{sl2solution1}) and (\ref{sl2solution2}).
\subsubsection*{The $(p,q)$ string.}
We have omitted to include the axion, $\chi$, in our discussion of the IIB theory bound states. The axion in the decomposition of $E_{11}$ to the IIB theory appears as the field associated to the step generator ${K^{\bar{1}}}_{\bar{2}}$ of $SL(2,\mathbb{R})$. It is the generator associated to node number ten as shown in figure \ref{E11IIB}. Using the notation of section 2.2 the associated simple root is
\begin{equation}
\alpha_{10}=g_1-g_2
\end{equation}
and its action on the Young tableaux of the IIB decomposition is to transform only the $SL(2,\mathbb{R})$ Young tableau. We recall that upon decomposition to the IIB theory, each root of $E_{11}$ is described by a pair of Young tableaux which are tensor of ($SL(10,\mathbb{R})$,$SL(2,\mathbb{R})$). The adjoint action of ${K^{\bar{1}}}_{\bar{2}}$ is trivial on generators whose $SL(2,\mathbb{R})$ tableau has columns all of height two (e.g. those associated to the D3, D7 or KK5 branes in table \ref{table:IIB}). But its action on tableaux with a single block labelled $\bar{2}$ is to lower them to $\bar{1}$, for example under the adjoint action of ${K^{\bar{1}}}_{\bar{2}}$ the F1 string is mapped to the D1 brane, together they form an $SL(2,\mathbb{R})$ doublet:
\begin{align}
\ytableausetup
{mathmode, boxsize=1.0em}
 & \left( \, \begin{ytableau}
\mbox{{\scriptsize 10}}\\
\mbox{{\scriptsize 9}}  \\
\end{ytableau} \, ,
\begin{ytableau}
\mbox{{\scriptsize $\bar{2}$}}  \\
\end{ytableau}  \, \right) \overset{{K^{\bar{1}}}_{\bar{2}}}{\xrightarrow{\hspace*{40pt}} } 
 \left( \, \begin{ytableau}
\mbox{{\scriptsize 10}}  \\
\mbox{{\scriptsize 9}}  \\
\end{ytableau} \, ,
\begin{ytableau}
\mbox{{\scriptsize $\bar{1}$}}  \\
\end{ytableau}  \, \right).\\
\nonumber &\hspace{20pt}  F1 \hspace{85pt} D1
\end{align}
We are using the barred numbers ($\bar{1}$,$\bar{2}$) to indicate the labels of the $SL(2,\mathbb{R})$ tensors. The generator ${K^{\bar{1}}}_{\bar{2}}$ is the analogue of the spacetime transformations ${K^i}_j$ acting on only the $SL(2,\mathbb{R})$ index of the IIB generator. The adjoint action in the $SL(2,\mathbb{R})$ sector of the theory as opposed to the $SL(10,\mathbb{R})$  spacetime sector indicates a new type of bound state but one which is equally well described as a null geodesic on the coset $\frac{SL(3,\mathbb{R})}{SO(1,2)}$. We will relate the solution we find from this example to the $(p,q)$ string \cite{Schwarz:1995p2725} which exhibits an $SL(2,\mathbb{Z})$ symmetry upon quantisation.

A representative coset element $g$ and solution fields are identical to those given in equations (\ref{sl2solution1}) and (\ref{sl2solution2}) which were originally found in \cite{Houart:2010p2928}.
The embedding of the generators which form the Borel sub-algebra of $\mathfrak{sl}(3,\mathbb{R})$ into $\mathfrak{e}_{11}$ do differ and in this example are 
\begin{align}
\nonumber H_1&=-\frac{1}{4}({K^1}_1+\ldots +{K^8}_8)+\frac{3}{4}({K^9}_9+{K^{10}}_{10})+{\hat{R}} ,\\
H_2&=-2{\hat{R}},\\
\nonumber E_1&=R^{910(\bar{2})}, \quad E_2={K^{\bar{1}}}_{\bar{2}} \quad \mbox{and} \quad E_{12}=-R^{910(\bar{1})}.
\end{align}
The metric is read from the solution to be
\begin{equation}
ds^2=N_1^{\frac{1}{4}}((dx^1)^2+(dx^2)^2+\ldots + (dx^8)^2+N_1^{-1}((dx^9)^2-(dt^{10})^2))
\end{equation}
and the dilaton is
\begin{equation}
e^{\hat{A}}=N_1^{-\frac{1}{2}}N_2
\end{equation}
where $N_1=1+\kappa\xi$, $N_2=1+\kappa\xi\cos^2{\beta}$ and $\kappa\in\mathbb{R}$ is a constant. The parameter $\beta\in [0,\frac{\pi}{2}]$ does not change the spacetime metric but only varies the dilaton and interpolates between the F1 solution when $\beta=0$ and the D1 solution when $\beta=\frac{\pi}{2}$.    We will discuss the constant $\kappa$ momentarily. First we derive the active gauge fields in the solution, we have,
\begin{align}
P_1&=\frac{\sqrt{\partial N_1 \alpha_{12}}}{N_1\sqrt{N_2}}, \qquad P_2=\frac{\sqrt{\partial N_2 \alpha_{12}}}{N_2 \sqrt{N_1}} \quad \mbox{and}\\
&\hspace{35pt} P_{12}=-\sqrt{\frac{\partial N_1 \partial N_2}{N_1N_2}}
\end{align}
where $\alpha_{12}\equiv N_2\partial N_1 -N_1\partial N_2=\kappa\sin^2\beta$. To determine the field strengths we use formula $F_i=(e^{\hat{A}})^{a_i} P_i$ where $[\hat{R},E_i]=a_i E_i$. Noting that,
\begin{equation}
a_1=\frac{1}{2}, \quad a_2=-1\quad \mbox{and}\quad a_{12}=-\frac{1}{2}
\end{equation}
we find
\begin{align}
H_{1910}&=-\partial N_1^{-1} \sin{\beta}+\frac{\partial N_1 \cos{\beta}}{N_1N_2}\\
\chi&=-\partial N_2^{-1}  \tan{\beta}
\end{align}
where we have identified $\xi=x^1$ and $\chi$ is the axion.

The $(p,q)$ string of \cite{Schwarz:1995p2725} is quantised due to the presence of the magnetically charged five-branes. We have not considered the effect of the presence of fivebranes in constructing our solution above which one may rescale and interpret as an interpolation between a (1,0) string and a (0,1) string. Our interpolation parameter is smooth and is a consequence of the action of the compact generator of the local sub-group $SO(1,2)$. However we may compare the solution of \cite{Schwarz:1995p2725} to enlarge our $SO(2)$ symmetry to $SL(2,\mathbb{Z})$ which will inform the quantisation procedure for $E_{11}$ and other bound state solutions of a similar type. To compare the fields of the interpolating solution with those of the $(p,q)$ string we note that $\beta=\theta+\frac{\pi}{2}$ and $N=A_q$ in the notation of \cite{Schwarz:1995p2725}. Consequently one finds that, after quantisation, $\beta$ may take only take a discrete values characterised by the co-prime integers $q$ and $p$ such that:
\begin{equation}
\cos{\beta}=\frac{p}{\sqrt{q^2+p^2}} \qquad \mbox{ and } \qquad \sin{\beta}=\frac{q}{\sqrt{q^2+p^2}}.
\end{equation}
We also fix the parameter $\kappa = Q\sqrt{q^2+p^2}$ and hence we identify the two harmonic functions in our description:
\begin{equation}
N_1=1+\sqrt{q^2+p^2}\frac{Q}{r^6} \qquad \mbox{ and } \qquad N_2=1+\frac{p^2}{\sqrt{q^2+p^2}}\frac{Q}{r^6}
\end{equation}
where we have unsmeared the solution so that $N_1$ and $N_2$ are harmonic functions in $r=\sqrt{(x^1)^2+(x^2)^2+\ldots +(x^8)^2}$. 

One can see that the transition from the coset model solution to the $(p,q)$ string has, in some sense, both generalised and restricted the initial solution which interpolated smoothly between a $(1,0)$ string and a $(0,1)$ string. In contrast the $(p,q)$ string solution jumps discretely between all possible $(p,q)$ strings, where $q$ and $p$ are co-prime. One may imagine that future developments of the coset model approach will be able to derive precisely the $(p,q)$ string and that the quantisation may be understood within a general framework. Work in this direction will be informed by counting the number of constants describing the solution. 

Herein we will not consider bound states with an axion field, but it is worth noting that there are similar $(p,q)$ bound states for the dual case of the bound state of D5 and NS5 branes. As the dimension of the $SL(2,\mathbb{R})$ representation that the gauge field transforms under increases the bound states become more complicated, in particular bound states of the D7 branes and the D9 branes with an active axion ought to be further investigated.
\end{subsubsection}

\begin{subsubsection}{Type-IIB: Cosets of rank three groups.}
As for the IIA examples we will list the bound states of D-branes without exotic content which have global 
$SL(4,\mathbb{R})$ and $SL(3,\mathbb{R})^+$ symmetries. 
There are 532 IIB bound states which possess an $SL(4,\mathbb{R})$ symmetry, of these only twelve
 bound states may be interpreted as pure bound states of branes, these cases are indicated in table \ref{IIB_rank_3_bound_states_finite}. In addition there are eleven bound states corresponding to the embedding of $SL(3,\mathbb{R})^+$ in $E_{11}$ listed in table  \ref{IIB_rank_3_bound_states_affine} and we will not discuss these examples further here.
\begin{table}[h!]
\centering
\begin{tabular}{ c || c | c }
& S1$_a$ & S1$_b$   \\
\hline
\hline
(F1,D3) & (F1,D3,D5) & (F1,D3,D3,NS5) \\
\hline
(F1,D5)  & (F1,D5,D7$_a$) & -  \\
\hline
(F1,D7$_a$) & (F1,D7$_a$,D9$_a$) & - \\
\hline
(D1,D3) & (D1,D3,D3,D5) & (D1,D3,NS5) \\
\hline
(D1,NS5) & - & (D1,NS5,D7$_b$)  \\
\hline
(D1,D7$_b$) & - & (D1,D7$_b$,D9$_b$)  \\
\hline
(D3,D5) & (D3,D5,D5,D7$_a$) & -  \\
\hline
(D3,NS5) & - & (D3,NS5,NS5,D7$_b$)  \\
\hline
(D5,D7$_a$) & (D5,D7$_a$,D7$_a$,D9$_a$) & -  \\
\hline
(NS5,D7$_b$) & - & (NS5,D7$_b$,D7$_b$,D9$_b$)  \\
\end{tabular}
\caption{The $SL(4,\mathbb{R})$ bound states of the canonical IIB branes.} \label{IIB_rank_3_bound_states_finite}
\end{table}
\begin{table}[h!]
\small
\tabcolsep 1pt
\centering
\begin{tabular}{ c || c | c | c | c | c | c }
& S1$_b$ & S3 & S5$_a$ & S5$_b$ & S7$_a$ & S7$_b$\\
\hline
\hline
(F1,D3) & (F1,D3,D5) & - & (F1,D3,D7$_a$) & - & (F1,D3,D9$_a$) & - \\
\hline
(F1,D5) & - & (F1,D5,D5) & (F1,D5,D7$_a$) & - & - & -\\
\hline
(D1,D3) & - & (F1,D3,NS5) & - & (D1,D3,D7$_b$) & - & (D1,D3,D9$_b$) \\
\hline
(D1,NS5)  & - & (D1,NS5,NS5) & - & (D1,NS5,D7$_b$) & - & - \\
\hline
(D3,D5)  & (D3,D5,NS5) & - & - & - & - & - \\
\end{tabular}
\caption{The $SL(3,\mathbb{R})^+$ bound states of the canonical IIB branes.} \label{IIB_rank_3_bound_states_affine}
\end{table}
\end{subsubsection}
\begin{subsubsection}{Type-IIB: Cosets of rank four groups.}
In the IIB decomposition there are no bound states of D-branes described by cosets of $SL(5,\mathbb{R})$. There are eight states which are cosets of $SO(4,4)$ and the orientations of the branes involved are shown in tables \ref{IIB_rank_4_so_8_1a}-\ref{IIB_rank_4_so_8_4b}. The bound states (F1,D3,D5$^2$,D7$_a$,KK4$_{2a}$), (D1,D3,NS5$^2$,D7$_b$,KK4$_{2b}$), (F1,D5,D7$_a^2$,D9$_a$,KK5$_{4a}$) and (D1,NS5,D7$_b^2$,D9$_b$,KK5$_{4b}$) are described by null geodesic motion on $\frac{SO(4,4)}{SO(1,3)\times SO(1,3)}$ while the bound states (D1,D3$^3$,D5$^3$,D7$_a$), (F1,D3$^3$,NS5$^3$,D7$_b$),   (D3,D5$^3$,D7$_a$$^3$,D9$_a$) and (D3,NS5$^3$,D7$_b$$^3$,D9$_b$) are given by the null geodesic on $\frac{SO(4,4)}{SO(2,2)\times SO(2,2)}$. We note that these eight bound states may be arranged into four S-dual pairs, for example (F1,D3,D5$^2$,D7$_a$,KK4$_{2a}$) and (D1,D3,NS5$^2$,D7$_b$,KK4$_{2b}$) are S-dual to each other.
\begin{table}[h!] 
\centering 
\scalebox {0.95} {
\begin{tabular}{ c | c | c | c | c | c | c | c | c | c | c  } 
Branes &1& 2& 3& 4& 5& 6& 7& 8& 9& 10(t)  \\
\hline
F1&&&&&&&&&$\bullet$&$\bullet$   \\
\hline 
D3&&&&&&&$\bullet$&$\bullet$ &$\bullet$ &$\bullet$   \\
\hline 
D5&&&&&&$\bullet$ &$\bullet$ &$\bullet$&$\bullet$&$\bullet$   \\
\hline 
D5&&&$\bullet$&$\bullet$&&&$\bullet$&$\bullet$&$\bullet$&$\bullet$   \\
\hline 
D7$_a$&&&$\bullet$&$\bullet$&$\bullet$&$\bullet$ &$\bullet$ &$\bullet$ &$\bullet$ &$\bullet$   \\
\hline 
KK4$_{2a}$&&&$\bullet$&$\bullet$&$\bullet$&$\bullet$ &$\odot$ &$\odot$ &$\bullet$ &$\bullet$   \\
\end{tabular} }
\caption{The $SO(4,4)$ bound state of (F1,D3,D5$^2$,D7$_a$,KK4$_{2a}$)}\label{IIB_rank_4_so_8_1a}
\end{table}
\begin{table}[h!] 
\centering 
\scalebox {0.95} {
\begin{tabular}{ c | c | c | c | c | c | c | c | c | c | c  } 
Branes &1& 2& 3& 4& 5& 6& 7& 8& 9& 10(t)  \\
\hline
D1&&&&&&&&&$\bullet$&$\bullet$   \\
\hline 
D3&&&&&&&$\bullet$&$\bullet$ &$\bullet$ &$\bullet$   \\
\hline 
NS5&&&&&&$\bullet$ &$\bullet$ &$\bullet$&$\bullet$&$\bullet$   \\
\hline 
NS5&&&$\bullet$&$\bullet$&&&$\bullet$&$\bullet$&$\bullet$&$\bullet$   \\
\hline 
D7$_b$&&&$\bullet$&$\bullet$&$\bullet$&$\bullet$ &$\bullet$ &$\bullet$ &$\bullet$ &$\bullet$   \\
\hline 
KK4$_{2b}$&&&$\bullet$&$\bullet$&$\bullet$&$\bullet$ &$\odot$ &$\odot$ &$\bullet$ &$\bullet$   \\
\end{tabular} }
\caption{The $SO(4,4)$ bound state of (D1,D3,NS5$^2$,D7$_b$,KK4$_{2b}$)}\label{IIB_rank_4_so_8_1b}
\end{table}
\begin{table}[h!] 
\centering 
\scalebox {0.95} {
\begin{tabular}{ c | c | c | c | c | c | c | c | c | c | c  } 
Branes &1& 2& 3& 4& 5& 6& 7& 8& 9& 10(t)  \\
\hline
F1&&&&&&&&&$\bullet$ &$\bullet$   \\
\hline 
D3&&&&&&&$\bullet$ &$\bullet$ &$\bullet$ &$\bullet$   \\
\hline 
D3&&&& &$\bullet$ &$\bullet$ & & &$\bullet$ &$\bullet$   \\
\hline 
D3&&&$\bullet$&$\bullet$ & & & & &$\bullet$ &$\bullet$   \\
\hline 
NS5&& &&&$\bullet$&$\bullet$ &$\bullet$ &$\bullet$ &$\bullet$ &$\bullet$   \\
\hline 
NS5&& &$\bullet$ &$\bullet$ &&&$\bullet$ &$\bullet$ &$\bullet$ &$\bullet$   \\
\hline 
NS5&& &$\bullet$ &$\bullet$ &$\bullet$&$\bullet$ & & &$\bullet$ &$\bullet$   \\
\hline 
D7$_b$&& &$\bullet$ &$\bullet$ &$\bullet$&$\bullet$ &$\bullet$ &$\bullet$ &$\bullet$ &$\bullet$   \\
\end{tabular} }
\caption{The $SO(4,4)$ bound states of (F1,D3$^3$,NS5$^3$,D7$_b$)}\label{IIB_rank_4_so_8_2a}
\end{table}
\begin{table}[h!] 
\centering 
\scalebox {0.95} {
\begin{tabular}{ c | c | c | c | c | c | c | c | c | c | c  } 
Branes &1& 2& 3& 4& 5& 6& 7& 8& 9& 10(t)  \\
\hline
D1&&&&&&&&&$\bullet$ &$\bullet$   \\
\hline 
D3&&&&&&&$\bullet$ &$\bullet$ &$\bullet$ &$\bullet$   \\
\hline 
D3&&&& &$\bullet$ &$\bullet$ & & &$\bullet$ &$\bullet$   \\
\hline 
D3&&&$\bullet$&$\bullet$ & & & & &$\bullet$ &$\bullet$   \\
\hline 
D5&& &&&$\bullet$&$\bullet$ &$\bullet$ &$\bullet$ &$\bullet$ &$\bullet$   \\
\hline 
D5&& &$\bullet$ &$\bullet$ &&&$\bullet$ &$\bullet$ &$\bullet$ &$\bullet$   \\
\hline 
D5&& &$\bullet$ &$\bullet$ &$\bullet$&$\bullet$ & & &$\bullet$ &$\bullet$   \\
\hline 
D7$_a$&& &$\bullet$ &$\bullet$ &$\bullet$&$\bullet$ &$\bullet$ &$\bullet$ &$\bullet$ &$\bullet$   \\
\end{tabular} }
\caption{The $SO(4,4)$ bound states of (D1,D3$^3$,D5$^3$,D7$_a$)}\label{IIB_rank_4_so_8_2b}
\end{table}
\begin{table}[h!] 
\centering 
\scalebox {0.95} {
\begin{tabular}{ c | c | c | c | c | c | c | c | c | c | c  } 
Branes &1& 2& 3& 4& 5& 6& 7& 8& 9& 10(t)  \\
\hline
F1&&&&&&&&&$\bullet$ &$\bullet$   \\
\hline 
D5&&&&&$\bullet$&$\bullet$&$\bullet$&$\bullet$&$\bullet$ &$\bullet$   \\
\hline 
D7$_a$&&&$\bullet$&$\bullet$&$\bullet$&$\bullet$ &$\bullet$ &$\bullet$ &$\bullet$ &$\bullet$   \\
\hline 
D7$_a$&$\bullet$&$\bullet$&&&$\bullet$ &$\bullet$&$\bullet$&$\bullet$ &$\bullet$ &$\bullet$   \\
\hline 
D9$_a$&$\bullet$&$\bullet$&$\bullet$&$\bullet$&$\bullet$ &$\bullet$ &$\bullet$ &$\bullet$ &$\bullet$&$\bullet$   \\
\hline 
KK5$_{4a}$&$\bullet$&$\bullet$&$\bullet$&$\bullet$&$\odot$ &$\odot$ &$\odot$ &$\odot$ &$\bullet$&$\bullet$   \\
\end{tabular} }
\caption{The $SO(4,4)$ bound states of (F1,D5,D7$_a^2$,D9$_a$,KK5$_{4a}$)}\label{IIB_rank_4_so_8_3a}
\end{table}
\begin{table}[h!] 
\centering 
\scalebox {0.95} {
\begin{tabular}{ c | c | c | c | c | c | c | c | c | c | c  } 
Branes &1& 2& 3& 4& 5& 6& 7& 8& 9& 10(t)  \\
\hline
D1&&&&&&&&&$\bullet$ &$\bullet$   \\
\hline 
NS5&&&&&$\bullet$&$\bullet$&$\bullet$&$\bullet$&$\bullet$ &$\bullet$   \\
\hline 
D7$_b$&&&$\bullet$&$\bullet$&$\bullet$&$\bullet$ &$\bullet$ &$\bullet$ &$\bullet$ &$\bullet$   \\
\hline 
D7$_b$&$\bullet$&$\bullet$&&&$\bullet$ &$\bullet$&$\bullet$&$\bullet$ &$\bullet$ &$\bullet$   \\
\hline 
D9$_b$&$\bullet$&$\bullet$&$\bullet$&$\bullet$&$\bullet$ &$\bullet$ &$\bullet$ &$\bullet$ &$\bullet$&$\bullet$   \\
\hline 
KK5$_{4b}$&$\bullet$&$\bullet$&$\bullet$&$\bullet$&$\odot$ &$\odot$ &$\odot$ &$\odot$ &$\bullet$&$\bullet$   \\
\end{tabular} }
\caption{The $SO(4,4)$ bound states of (D1,NS5,D7$_b^2$,D9$_b$,KK5$_{4b}$)}\label{IIB_rank_4_so_8_3b}
\end{table}
\begin{table}[h!] 
\centering 
\scalebox {0.95} {
\begin{tabular}{ c | c | c | c | c | c | c | c | c | c | c  } 
Branes &1& 2& 3& 4& 5& 6& 7& 8& 9& 10(t)  \\
\hline
D3&&&&&&&$\bullet$ &$\bullet$ &$\bullet$ &$\bullet$   \\
\hline 
D5&&&&&$\bullet$ &$\bullet$ &$\bullet$ &$\bullet$ &$\bullet$ &$\bullet$   \\
\hline 
D5&&&$\bullet$ &$\bullet$ & &&$\bullet$ &$\bullet$ &$\bullet$ &$\bullet$   \\
\hline 
D5&$\bullet$ &$\bullet$ & &&&&$\bullet$ &$\bullet$ &$\bullet$ &$\bullet$   \\
\hline 
D7$_a$&&&$\bullet$ &$\bullet$ &$\bullet$ &$\bullet$ &$\bullet$ &$\bullet$ &$\bullet$ &$\bullet$   \\
\hline 
D7$_a$&&$\bullet$ &$\bullet$ &&&$\bullet$ &$\bullet$ &$\bullet$ &$\bullet$ &$\bullet$   \\
\hline 
D7$_a$&$\bullet$ &$\bullet$ && &$\bullet$ &$\bullet$ &$\bullet$ &$\bullet$ &$\bullet$ &$\bullet$   \\
\hline 
D9$_a$&$\bullet$ &$\bullet$ &$\bullet$ &$\bullet$ &$\bullet$ &$\bullet$ &$\bullet$ &$\bullet$ &$\bullet$ &$\bullet$   \\
\end{tabular} }
\caption{The $SO(4,4)$ bound states of (D3,D5$^3$,D7$_a^3$,D9$_a$)}\label{IIB_rank_4_so_8_4a}
\end{table}
\begin{table}[h!] 
\centering 
\scalebox {0.95} {
\begin{tabular}{ c | c | c | c | c | c | c | c | c | c | c  } 
Branes &1& 2& 3& 4& 5& 6& 7& 8& 9& 10(t)  \\
\hline
D3&&&&&&&$\bullet$ &$\bullet$ &$\bullet$ &$\bullet$   \\
\hline 
NS5&&&&&$\bullet$ &$\bullet$ &$\bullet$ &$\bullet$ &$\bullet$ &$\bullet$   \\
\hline 
NS5&&&$\bullet$ &$\bullet$ & &&$\bullet$ &$\bullet$ &$\bullet$ &$\bullet$   \\
\hline 
NS5&$\bullet$ &$\bullet$ & &&&&$\bullet$ &$\bullet$ &$\bullet$ &$\bullet$   \\
\hline 
D7$_b$&&&$\bullet$ &$\bullet$ &$\bullet$ &$\bullet$ &$\bullet$ &$\bullet$ &$\bullet$ &$\bullet$   \\
\hline 
D7$_b$&&$\bullet$ &$\bullet$ &&&$\bullet$ &$\bullet$ &$\bullet$ &$\bullet$ &$\bullet$   \\
\hline 
D7$_b$&$\bullet$ &$\bullet$ && &$\bullet$ &$\bullet$ &$\bullet$ &$\bullet$ &$\bullet$ &$\bullet$   \\
\hline 
D9$_b$&$\bullet$ &$\bullet$ &$\bullet$ &$\bullet$ &$\bullet$ &$\bullet$ &$\bullet$ &$\bullet$ &$\bullet$ &$\bullet$   \\
\end{tabular} }
\caption{The $SO(4,4)$ bound states of (D3,NS5$^3$,D7$_b^3$,D9$_b$)}\label{IIB_rank_4_so_8_4b}
\end{table}

\end{subsubsection}
\end{subsection}
\begin{subsection}{Embedding of groups of rank five and above in $E_{11}$.}
In this section we join together the discussion of IIA and IIB solutions. The reader who is interested in the precise details of the bound states is encouraged to scour the accompanying catalogues to find the interesting examples we will discuss here.
\begin{subsubsection}{Exotic states from ranks 5 to 8}
There are no pure brane bound states beyond rank four. From ranks five to eight while all the usual $A_n$, $D_n$ and $E_n$ algebras appear they all contain within the bound state mixed symmetry tensors beyond the KK5 brane gauge field and are classed as exotic (as they contain at least one exotic brane) or non-geometric. Consequently in this section we will restrict ourselves to comment on some of the problems which occur. 

The most significant obstruction to the universal application of the coset model is that the majority of the solutions are associated to Dynkin diagrams of indefinite type. These solutions have not been included in the catalogue of results associated to this paper. We are optimistic that it will prove possible to recast the null geodesic motion on these indefinite cosets as a more complicated but solvable motion on a set of simple cosets. This problem would be best approached by gaining an understanding of the geodesic motion on affine cosets. The affine cosets present an odd situation as the embedded algebra is infinite, and there are consequently an infinite set of equations of motion that must be simultaneously solved. One may hope that work on applying constraint equations \cite{Damour:2007p623,Damour:2011p2920} may be adapted to these algebras to allow the controlled elimination of an infinite set of generators - it would be particularly useful to be able to find a constraint that eliminated all the null roots for example.

Another problem occurs for bound state solutions of exotic branes having recognisable Dynkin diagrams for which the bound state is space-time filling. The coset model which parameterises a null geodesic by $\xi$ must be embedded in space-time and the present solution ansatz takes $\xi$ to be tranverse to all the constituent branes. A trivial embedding of $\xi$ into space-time can then be achieved by identifying $\xi$ with one transverse space-time coordinate. For each of the symmetries $SL(6,\mathbb{R})$, $SL(7,\mathbb{R})$, $SL(8,\mathbb{R})$, $SL(9,\mathbb{R})$, $SO(5,5)$, $SO(6,6)$, $SO(7,7)$, $SO(8,8)$, $E_6$, $E_7$ and $E_8$ there are examples where there are no common transverse directions. One may hope that for some cases a nontrivial embedding of the null geodesic parameter on the coset may be possible. A non-trivial embedding will have an effect on the harmonic functions which appear in the solution. Alternatively one may be able to loosen the harmonic ansatz in some way when solving the equations of motion and avoid the utility of a totally transverse direction. It is worth noting that one can find many examples of non-space-time-filling bound states in all the cases mentioned above.

One may wonder how large a symmetry one can find constructed out of the basic brane roots. It may seem naively possible to construct symmetry groups of arbitrary rank, but as the roots are embedded in an eleven-dimensional vector space there will be a limit of rank eleven on the size of the associated Dynkin diagram. This does not mean that we will be able to construct such a rank eleven Dynkin diagram from canonical solutions and indeed the results find the largest sub-algebra that can be embedded using only the generators associated to the canonical branes is of rank ten.

There is a new class of solution within the ADE Dynkin diagrams for ranks greater than eight. $E_9$ is an affine symmetry and will possess a large part of the full $E_{11}$ symmetry encoded as a solution. We have found numerous rank nine symmetries corresponding to $A_9$, $D_9$ and $E_9$. At rank ten the possibilities are more limited but we do find bound states containing exotic branes having global $A_{10}$, $D_{10}$ and $E_{10}$ symmetry. These can all be looked up in the catalogue associated to this paper. We present in table \ref{rank_10_e_10} one example of the $E_{10}$ symmetry appearing in the IIA decomposition although there are others. All the states in the example, apart from an initial $D4$ brane, include exotic branes.
\begin{table}[ht] 
\centering 
\scalebox {0.95} {
\begin{tabular}{ c | c | c | c | c | c | c | c | c | c | c  } 
Brane roots &1& 2& 3& 4& 5& 6& 7& 8& 9& 10(t)  \\
\hline
D4&&&&&&$\bullet$&$\bullet$&$\bullet$&$\bullet$&$\bullet$  \\
\hline 
D4&&&$\bullet$&$\bullet$&$\bullet$&&&&$\bullet$&$\bullet$  \\
\hline 
D4&&$\bullet$&&&$\bullet$&&$\bullet$&$\bullet$&&$\bullet$  \\
\hline 
D4&$\bullet$&&&$\bullet$&&&&$\bullet$&$\bullet$&$\bullet$  \\
\hline 
D4&&&&$\bullet$&$\bullet$&$\bullet$&$\bullet$&&&$\bullet$  \\
\hline 
D4&&&$\bullet$&&$\bullet$&$\bullet$&&$\bullet$&&$\bullet$  \\
\hline 
D4&$\bullet$&&&&$\bullet$&&$\bullet$&&$\bullet$&$\bullet$  \\
\hline 
D6&&$\bullet$&&$\bullet$&$\bullet$&$\bullet$&&$\bullet$&$\bullet$&$\bullet$  \\
\hline 
D0&&&&&&&&&&$\bullet$  \\
\hline 
S2&&&&&$\bullet$&&&$\bullet$&$\bullet$&  \\
\end{tabular} }
\caption{An $E_{10}$ bound state of  a D4 brane and exotic branes.}\label{rank_10_e_10}
\end{table}
The bound state will be described by a null geodesic motion on the coset $\frac{E_{10}}{{\cal K}(E_{10})}$. This is the same description used to describe chaotic billiard motion in the vicinity of a cosmological singularity \cite{Damour:2002p617}
\end{subsubsection}
\end{subsection}
\end{section}

\begin{section}{Discussion}
The use of the motion of a massless particle on cosets to describe solutions from Kac-Moody algebras dates back to the analysis of cosmological singularities and $\mathfrak{e}_{10}$ in \cite{Damour:2002p617,Damour:2004p4836} and the brane $\sigma$-model of \cite{Englert:2004p2874,Englert:2005p966}. In the initial work considering the null geodesic motion the hyperbolic algebra $\mathfrak{e}_{10}$ was used to develop algebraically the chaotic billiard ball dynamics identified with the Weyl chamber of $\mathfrak{e}_{10}$ near a space-like singularity \cite{Damour:2001p3108} for a review see \cite{Damour:2003p620}.

The setting of $\mathfrak{e}_{10}$ lent itself naturally to consider a null geodesic motion parameterised only by time on $\frac{{\cal G}^{++}}{{\cal K}({\cal G}^{++})}$. The brane $\sigma$-model developed the model by considering a motion on a $\frac{{\cal G}^{+++}}{{\cal K}({\cal G}^{+++})}$ coset in which the fields of the theory incorporated a democracy between spatial and temporal coordinates. It was possible to use the machinery to successfully recover $\frac{1}{2}$-BPS brane solutions of M-theory, the precise solutions to the model having previously been presented in the form of a generic representative element of the coset \cite{West:2004p1593}. 

The coset model describing the chaotic billiard motion near a space-like singularity parameterised the evolution of the fields by time, in later models the parameter describing the motion was trivially embedded within space-time, transverse to the constituent brane world-volumes. The consideration of a lightlike particle motion on the coset remained sufficient to describe even the bound state solutions. In contrast to the vicinity of a space-like singularity where spatial degrees of freedom are suppressed, the small finite cosets used to describe bound states very readily possess more coordinates in the coset than can be trivially embedded in space-time. For example $\frac{SL(5,\mathbb{R})}{SO(2,3)}$ is fully parameterised by fourteen coordinates. If the model did not restrict our considerations to the null geodesic motion on the coset, one would think that coordinates had been generated which is in marked contrast to their suppression near a cosmological singularity. The use of a null geodesic allows one to generate brane solutions dependent upon one coordinate associated to space-time. One may wonder whether the model may be extended to include dependence upon a greater number of parameters while retaining a meaningful Lagrangian. One suggestion in \cite{Damour:2002p617} was to consider adding transcendental invariants of the Kac-Moody algebra to the Lagrangian and deserves further investigation.

In this paper we have explored exhaustively the embeddings of sub-groups $\cal G$ into $E_{11}$ which are associated to bound states of the canonical string theory branes. The results are vast and we have presented numerous challenges to the programme of finding bound state solutions associated to null geodesic motion on a general coset, principally the challenges of the affine coset and co-dimension zero branes. Additionally we have found new solutions identifying bound states to the cosets of $SL(5,\mathbb{R})$. The generalisation to find solutions on the cosets $SL(n,\mathbb{R})$ for $5<n<11$ is straightforward as the $P_i$ for the solution follow a simple pattern. However there remain challenges to understand solutions for real forms of the sub-algebra $SO(n)$ where the present ansatz is insufficient.

Our aim was to explore the realm of the possible and to seek out algebraic obstructions to interpreting mixed symmetry Young tableaux as bound states of string theory branes. We have focussed on the bound states containing only form fields or the gauge field for the KK5 monopole and have not yet found an impediment to continuing the analysis to include mixed-symmetry gauge fields in the bound states. While there may be problems for the cosets whose bound state solution is space-time-filling there are vast numbers of other simple cosets embedded in $\mathfrak{e}_{11}$ that also contain exotic branes and are solvable. Such bound states include those having straightforward equations of motion being described by cosets of $SL(n,\mathbb{R})$. It is anticipated that these bound states which include exotic branes will be relevant to the counting of black hole microstates \cite{Boer:2010p3099} and in section 3.2 we have included some rules for constructing stable objects including the exotic branes.

There remain a number of outstanding questions to be followed up. These include the supersymmetric properties of the bound state solution (see for example \cite{Houart:2011p4137} where a solution which breaks all supersymmetry is presented) and what happens to the preserved supercharges as the interpolation parameters are varied, there seems to be an inherent conflict between the continuous nature of the compact symmetries and the discrete change in the number of preserved supersymmetries. 
In the present work we have considered solutions associated to sets of real roots. A solution in \cite{Houart:2011p4137} showed it was possible to study solutions associated to the null roots of $E_{11}$, we may wish to open Pandora's box a little further and wonder if we can find solutions involving only imaginary roots. It would also be interesting to investigate the role of complex structure in the solutions and indeed to investigate more closely the unusual cosets $\frac{SO(4,4)}{SO(1,3)\times SO(1,3)}$ and $\frac{SO(4,4)}{SO(2,2)\times SO(2,2)}$ which have appeared unbidden in this work.

\end{section}
\section*{Acknowledgements}
It is a pleasure to thank Ling Bao, Nicolas Boulanger, Thibault Damour, Michael Fleming and Peter West for discussions during the course of this work. We thank the referee of our JHEP submission for detailed comment which have improved the clarity of the presentation. In addition I would like to thank the I.H.E.S., Bures-sur-Yvette, for providing a peaceful work environment and excellent hospitality for a month where a crucial part of this work was completed. Some of the Dynkin diagrams in this paper have been produced using Teake Nutma's excellent software SimpLie. This work has been supported by the STFC rolling grant (ST/G00395/1) of the theoretical physics group at King's College London.

\bibliographystyle{utphys}
\bibliography{stringtheoryboundstates}

\providecommand{\href}[2]{#2}\begingroup\raggedright\begin{thebibliography}{10}

\bibitem{West:2001p131}
P.~West, ``E11 and {M} theory,'' {\em Quantum and Classical Gravity} {\bf 18}
  (2001)  4443--4460, \href{http://arxiv.org/abs/hep-th/0104081v2}{{\tt
  hep-th/0104081v2}}.

\bibitem{Damour:2001p3108}
T.~Damour and M.~Henneaux, ``{E10}, {BE10} and arithmetical chaos in
  superstring cosmology,'' {\em Phys.Rev.Lett.} {\bf 86} (2001)  4749--4752,
  \href{http://arxiv.org/abs/hep-th/0012172v2}{{\tt hep-th/0012172v2}}.

\bibitem{Mizoguchi:1998p4868}
S.~Mizoguchi, ``{E10} symmetry in one-dimensional supergravity,'' {\em
  Nucl.Phys.B} {\bf 528} (1998)  238--264,
  \href{http://arxiv.org/abs/hep-th/9703160v4}{{\tt hep-th/9703160v4}}.

\bibitem{Julia:1981p4403}
B.~Julia, ``Group disintegrations,'' {\em in Superspace {\&} Supergravity, p.
  331, eds. S.W. Hawking and M. Roˇcek, Cambridge University Press.} (1981)  .

\bibitem{Nicolai:1992p4869}
H.~Nicolai, ``A hyperbolic {L}ie algebra from supergravity.,'' {\em Phys.Lett.
  B} {\bf 276} (1992)  333--340.

\bibitem{Kleinschmidt:2004p371}
A.~Kleinschmidt, I.~Schnakenburg, and P.~West, ``Very-extended {K}ac-{M}oody
  algebras and their interpretation at low levels,'' {\em Classical and Quantum
  Gravity} {\bf 21} (2004)  2493--2525,
  \href{http://arxiv.org/abs/hep-th/0309198v1}{{\tt hep-th/0309198v1}}.

\bibitem{Bergshoeff:2005p935}
E.~A. Bergshoeff, M.~de~Roo, S.~F. Kerstan, and F.~Riccioni, ``{IIB}
  supergravity revisited,'' {\em JHEP} {\bf 0508:098} (2005)  ,
  \href{http://arxiv.org/abs/hep-th/0506013v3}{{\tt hep-th/0506013v3}}.

\bibitem{Schnakenburg:2002p2747}
I.~Schnakenburg and P.~West, ``Massive {IIA} supergravity as a non-linear
  realisation,'' {\em Phys.Lett.B} {\bf 540} (2002)  137--145,
  \href{http://arxiv.org/abs/hep-th/0204207v1}{{\tt hep-th/0204207v1}}.

\bibitem{Riccioni:2007p609}
F.~Riccioni and P.~West, ``The {E(11)} origin of all maximal supergravities,''
  {\em JHEP} {\bf 0707:063} (2007)  ,
  \href{http://arxiv.org/abs/hep-th/0705.0752v1}{{\tt hep-th/0705.0752v1}}.

\bibitem{Riccioni:2009p1980}
F.~Riccioni, D.~Steele, and P.~West, ``The {E(11)} origin of all maximal
  supergravities - the hierarchy of field-strengths,'' {\em JHEP} {\bf
  0909:095} (2009)  , \href{http://arxiv.org/abs/0906.1177v1}{{\tt
  0906.1177v1}}.

\bibitem{Bergshoeff:2007p1300}
E.~Bergshoeff, I.~D. Baetselier, and T.~Nutma, ``E(11) and the embedding
  tensor,'' {\em JHEP} {\bf 0709:047} (2007)  ,
  \href{http://arxiv.org/abs/0705.1304v3}{{\tt 0705.1304v3}}.

\bibitem{Wit:2008p4770}
B.~de~Wit, H.~Nicolai, and H.~Samtleben, ``Gauged supergravities, tensor
  hierarchies, and m-theory,'' {\em JHEP} {\bf 0802:044} (2008)  ,
  \href{http://arxiv.org/abs/0801.1294v2}{{\tt 0801.1294v2}}.

\bibitem{Bergshoeff:2008p4773}
E.~Bergshoeff, J.~Gomis, T.~Nutma, and D.~Roest, ``Kac-moody spectrum of
  (half-)maximal supergravities,'' {\em JHEP} {\bf 0802:069} (2008)  ,
  \href{http://arxiv.org/abs/0711.2035v2}{{\tt 0711.2035v2}}.

\bibitem{Cook:2008p936}
P.~P. Cook and P.~West, ``Charge multiplets and masses for {E11},'' {\em JHEP}
  {\bf 0811:091} (2008)  , \href{http://arxiv.org/abs/0805.4451v2}{{\tt
  0805.4451v2}}.

\bibitem{Bergshoeff:2010p4753}
E.~A. Bergshoeff and F.~Riccioni, ``D-brane wess-zumino terms and u-duality,''
  {\em JHEP} {\bf 1011:139} (2010)  ,
  \href{http://arxiv.org/abs/1009.4657v2}{{\tt 1009.4657v2}}.

\bibitem{Bergshoeff:2011p4751}
E.~A. Bergshoeff and F.~Riccioni, ``String solitons and t-duality,'' {\em JHEP}
  {\bf 1105:131} (2011)  , \href{http://arxiv.org/abs/1102.0934v3}{{\tt
  1102.0934v3}}.

\bibitem{Bergshoeff:2011p4749}
E.~A. Bergshoeff and F.~Riccioni, ``Dual doubled geometry,'' {\em Phys.Lett. B}
  {\bf 702} (2011)  281--285, \href{http://arxiv.org/abs/1106.0212v1}{{\tt
  1106.0212v1}}.

\bibitem{Bergshoeff:2011p4758}
E.~A. Bergshoeff and F.~Riccioni, ``Branes and wrapping rules,'' {\em Physics
  Letters B} {\bf 704} (2011)  367--372,
  \href{http://arxiv.org/abs/1108.5067v1}{{\tt 1108.5067v1}}.

\bibitem{Bergshoeff:2011p4756}
E.~A. Bergshoeff and F.~Riccioni, ``The d-brane u-scan,'' {\em arXiv} {\bf
  hep-th} (2011)  , \href{http://arxiv.org/abs/1109.1725v1}{{\tt 1109.1725v1}}.

\bibitem{Bergshoeff:2012p4742}
E.~Bergshoeff, T.~Ortin, and F.~Riccioni, ``Defect branes,'' {\em Nuclear
  Physics B} {\bf 856} (2012)  210--227,
  \href{http://arxiv.org/abs/1109.4484v1}{{\tt 1109.4484v1}}.

\bibitem{Kleinschmidt:2011p4762}
A.~Kleinschmidt, ``Counting supersymmetric branes,'' {\em JHEP} {\bf 1110:144}
  (2011)  , \href{http://arxiv.org/abs/1109.2025v1}{{\tt 1109.2025v1}}.

\bibitem{Damour:2002p617}
T.~Damour, M.~Henneaux, and H.~Nicolai, ``E10 and a 'small tension expansion'
  of {M} theory,'' {\em Phys.Rev.Lett.} {\bf 89:221601} (2002)  ,
  \href{http://arxiv.org/abs/hep-th/0207267v1}{{\tt hep-th/0207267v1}}.

\bibitem{Riccioni:2006p600}
F.~Riccioni and P.~West, ``Dual fields and {E}11,'' {\em Physics Letters} {\bf
  B645} (2006)  286--292, \href{http://arxiv.org/abs/hep-th/0612001v1}{{\tt
  hep-th/0612001v1}}.

\bibitem{Geroch:1971p4776}
R.~Geroch, ``A method for generating solutions of einstein's equations,'' {\em
  Journal of Mathematical Physics} {\bf 12} (1971) no.~6, 918--924.

\bibitem{Geroch:1972p4777}
R.~Geroch, ``A method for generating new solutions of einstein's equation.
  ii,'' {\em Journal of Mathematical Physics} {\bf 13} (1972) no.~3, 394--404.

\bibitem{Breitenlohner:1987p4778}
P.~Breitenlohner and D.~Maison, ``On the {G}eroch group,'' {\em Annales
  Poincare Phys.Theor. 46 215} (1987)  .

\bibitem{Englert:2007p605}
F.~Englert, L.~Houart, A.~Kleinschmidt, H.~Nicolai, and N.~Tabti, ``An {E9}
  multiplet of {BPS} states,'' {\em Journal of High Energy Physics} {\bf
  0705:065} (2007)  , \href{http://arxiv.org/abs/hep-th/0703285v1}{{\tt
  hep-th/0703285v1}}.

\bibitem{Cook:2009p2751}
P.~P. Cook, ``Exotic {E}11 branes as composite gravitational solutions,'' {\em
  Classical and Quantum Gravity} {\bf 26} (2009) no.~253023, ,
  \href{http://arxiv.org/abs/0908.0485}{{\tt 0908.0485}}.

\bibitem{Houart:2010p2928}
L.~Houart, A.~Kleinschmidt, and J.~L. H{\"o}rnlund, ``Some algebraic aspects of
  half-{BPS} bound states in {M}-theory,'' {\em JHEP} {\bf 1003 :022} (2010)  ,
  \href{http://arxiv.org/abs/0911.5141v1}{{\tt 0911.5141v1}}.

\bibitem{West:2004p1593}
P.~West, ``The {IIA}, {IIB} and eleven dimensional theories and their common
  {E11} origin,'' {\em Nuclear Physics} {\bf B693} (2004)  76--102,
  \href{http://arxiv.org/abs/hep-th/0402140v2}{{\tt hep-th/0402140v2}}.

\bibitem{Englert:2004p2874}
F.~Englert and L.~Houart, ``G+++ invariant formulation of gravity and
  {M}-theories: {E}xact {BPS} solutions,'' {\em JHEP} {\bf 0401:002} (2004)  ,
  \href{http://arxiv.org/abs/hep-th/0311255v2}{{\tt hep-th/0311255v2}}.

\bibitem{Izquierdo:1995p1636}
J.~M. Izquierdo, N.~D. Lambert, G.~Papadopoulos, and P.~K. Townsend, ``Dyonic
  membranes,'' {\em Nuclear Physics} {\bf B460} (1995)  560--578.

\bibitem{Schnakenburg:2001p2748}
I.~Schnakenburg and P.~West, ``{K}ac-{M}oody symmetries of {IIB}
  supergravity,'' {\em Phys.Lett. B} {\bf 517} (2001)  421--428,
  \href{http://arxiv.org/abs/hep-th/0107181v1}{{\tt hep-th/0107181v1}}.

\bibitem{Tseytlin:1996p1683}
A.~A. Tseytlin, ``No-force condition and {BPS} combinations of p-branes in 11
  and 10 dimensions,'' {\em Nuclear Physics} {\bf B487} (1996)  141--154,
  \href{http://arxiv.org/abs/hep-th/9609212v2}{{\tt hep-th/9609212v2}}.

\bibitem{Tseytlin:1996p1469}
A.~A. Tseytlin, ``Harmonic superpositions of {M}-branes,'' {\em Nuclear
  Physics} {\bf B475} (1996)  149--163,
  \href{http://arxiv.org/abs/hep-th/9604035v4}{{\tt hep-th/9604035v4}}.

\bibitem{Argurio:1997p1652}
R.~Argurio, F.~Englert, and L.~Houart, ``Intersection rules for p-branes,''
  {\em Physics Letters} {\bf B398} (1997)  61--68,
  \href{http://arxiv.org/abs/hep-th/9701042v2}{{\tt hep-th/9701042v2}}.

\bibitem{Englert:2004p2875}
F.~Englert and L.~Houart, ``G+++ invariant formulation of gravity and
  {M}-theories: Exact intersecting brane solutions,'' {\em JHEP} {\bf 0405:059}
  (2004)  , \href{http://arxiv.org/abs/hep-th/0405082v1}{{\tt
  hep-th/0405082v1}}.

\bibitem{Bergshoeff:1997p1476}
E.~Bergshoeff, M.~de~Roo, E.~Eyras, B.~Janssen, and J.~P. van~der Schaar,
  ``Multiple intersections of {D}-branes and {M}-branes,'' {\em Nucl.Phys.B}
  {\bf 494} (1997)  119--143, \href{http://arxiv.org/abs/hep-th/9612095v3}{{\tt
  hep-th/9612095v3}}.

\bibitem{Mateos:2001p3100}
D.~Mateos and P.~K. Townsend, ``Supertubes,'' {\em Phys.Rev.Lett.} {\bf 87
  011602} (2001)  , \href{http://arxiv.org/abs/hep-th/0103030v2}{{\tt
  hep-th/0103030v2}}.

\bibitem{Larsson:2001p2275}
H.~Larsson, ``A note on half-supersymmetric bound states in {M}-theory and type
  {IIA},'' {\em arXiv} {\bf hep-th} (2001)  ,
  \href{http://arxiv.org/abs/hep-th/0105083v2}{{\tt hep-th/0105083v2}}.

\bibitem{Schwarz:1995p2725}
J.~H. Schwarz, ``An {SL(2,Z)} multiplet of type {IIB} superstrings,'' {\em
  Physics Letters} {\bf B360} (1995)  13--18,
  \href{http://arxiv.org/abs/hep-th/9508143v5}{{\tt hep-th/9508143v5}}.

\bibitem{Damour:2007p623}
T.~Damour, A.~Kleinschmidt, and H.~Nicolai, ``Constraints and the {E10} coset
  model,'' {\em Class.Quant.Grav.} {\bf 24} (2007)  6097--6120,
  \href{http://arxiv.org/abs/0709.2691v1}{{\tt 0709.2691v1}}.

\bibitem{Damour:2011p2920}
T.~Damour, A.~Kleinschmidt, and H.~Nicolai, ``Sugawara-type constraints in
  hyperbolic coset models,'' {\em Commun.Math.Phys.} {\bf 302} (2011)
  755--788, \href{http://arxiv.org/abs/0912.3491v1}{{\tt 0912.3491v1}}.

\bibitem{Damour:2004p4836}
T.~Damour and H.~Nicolai, ``Eleven dimensional supergravity and the e10/ke10
  sigma-model at low a9 levels,'' {\em Invited talk at 25th International
  Colloquium on Group Theoretical Methods in Physics (ICGTMP 2004), Cocoyoc,
  Mexico, 2-6 Aug 2004.} (2004)  ,
  \href{http://arxiv.org/abs/hep-th/0410245v2}{{\tt hep-th/0410245v2}}.

\bibitem{Englert:2005p966}
F.~Englert, M.~Henneaux, and L.~Houart, ``From very-extended to overextended
  gravity and {M}-theories,'' {\em JHEP} {\bf 0502:070} (2005)  ,
  \href{http://arxiv.org/abs/hep-th/0412184v1}{{\tt hep-th/0412184v1}}.

\bibitem{Damour:2003p620}
T.~Damour, M.~Henneaux, and H.~Nicolai, ``Cosmological billiards,'' {\em
  Class.Quant.Grav.} {\bf 20} (2003)  R145--R200,
  \href{http://arxiv.org/abs/hep-th/0212256v1}{{\tt hep-th/0212256v1}}.

\bibitem{Boer:2010p3099}
J.~de~Boer and M.~Shigemori, ``Exotic branes and non-geometric backgrounds,''
  {\em Phys.Rev.Lett.} {\bf 104 251603} (2010)  ,
  \href{http://arxiv.org/abs/1004.2521v1}{{\tt 1004.2521v1}}.

\bibitem{Houart:2011p4137}
L.~Houart, A.~Kleinschmidt, and J.~L. H{\"o}rnlund, ``An {M}-theory solution
  from null roots in {E}11,'' {\em JHEP} {\bf 1101:154} (2011)  ,
  \href{http://arxiv.org/abs/1101.2816v1}{{\tt 1101.2816v1}}.

\end{thebibliography}\endgroup

\end{document}